\documentclass[10pt]{IEEEtran}
\makeatletter
\def\ps@headings{%
\def\@oddhead{\mbox{}\scriptsize\rightmark \hfil \thepage}%
\def\@evenhead{\scriptsize\thepage \hfil \leftmark\mbox{}}%
\def\@oddfoot{}%
\def\@evenfoot{}}
\makeatother 
\usepackage{amssymb,amsmath}
\usepackage{color,soul}
\usepackage{amsfonts}
\usepackage{array}
\usepackage{algorithm,algorithmic}
\usepackage[nodisplayskipstretch]{setspace}
\usepackage{mathrsfs}
\usepackage{graphicx}
\usepackage{multirow}
\usepackage{epstopdf}
\usepackage{epsfig}
\usepackage{stmaryrd}
\usepackage{multirow,url}
\usepackage{color,}
\usepackage{amsthm}
\usepackage{bm}
\usepackage{pgf,tikz}
\usepackage{subfigure,cite}

\usepackage{fancyhdr,ifthen}

\newcounter{section:outage-analysis}
\begin{document}

\title{Optimizing Data Forwarding from Body Area Networks in the Presence of Body Shadowing with Dual Wireless Technology Nodes}

\author{Antonios Argyriou,~\IEEEmembership{Member,~IEEE}, Alberto Caballero Breva, and Marc Aoun\thanks{Antonios Argyriou is with the Department of Electrical and Computer Engineering, University of Thessaly, Greece (anargyr@ieee.org), Alberto Caballero Breva is with the University of Seville, Spain, Marc Aoun is with Philips Research, Eindhoven, The Netherlands.}
 \vspace{-6mm} }

\maketitle


\begin{abstract}
In this paper we are concerned with the problem of data forwarding from a wireless body area network (WBAN) to a gateway when body shadowing affects the ability of WBAN nodes to communicate with the gateway. To solve this problem we present a new WBAN architecture that uses two communication technologies. One network is formed between on-body nodes, and is realized with capacitive body-coupled communication (BCC), while an IEEE 802.15.4 radio frequency (RF) network is used for forwarding data to the gateway. WBAN nodes that have blocked RF links due to body shadowing forward their data through the BCC link to a node that acts as a relay and has an active RF connection. For this architecture we design a network layer protocol that manages the two communication technologies and is responsible for relay selection and data forwarding. Next, we develop analytical performance models of the medium access control (MAC) protocols of the two independent communication links in order to be used for driving the decisions of the previous algorithms. Finally, the analytical models are used for further optimizing energy and delay efficiency. We test our system under different configurations first by performing simulations and next by using real RF traces.
\end{abstract}


\section{Introduction}
\begin{figure}[t]
\begin{center}
  \includegraphics[keepaspectratio,width = 0.75\linewidth]{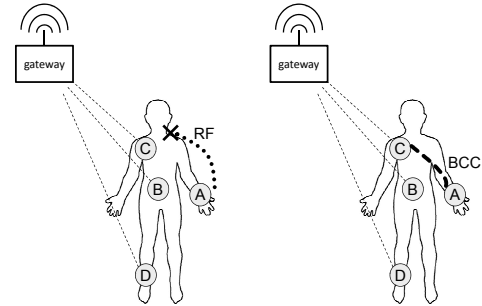}
  \caption{Nodes in a WBAN maybe unable to communicate to a gateway due to body shadowing (left). In the proposed system the BCC link is used for forwarding data to the nodes that have the optimal RF link (right).}
  \label{fig:dual-rf-bcc-principle}
\end{center}
\end{figure}

\IEEEPARstart{I}{n} several applications where wireless body area networks (WBAN) are deployed around the human body, reliable and low delay communication is of paramount importance because of the critical nature of the collected data (e.g. heart rate, blood pressure). Energy consumption is also key for the prolonged operation of the devices attached to the human body. For their communication needs these WBANs usually employ radio frequency (RF) technologies that operate in the industrial, scientific and medical (ISM) radio band. One of the dominant solutions is the IEEE 802.15.4 standard that is engineered specifically for low power devices~\cite{IEEE-802154,survey-wban11}. However, it is also possible to use the widely popular wireless LAN (WLAN) standard IEEE 802.11~\cite{IEEE-80211,cavalcanti07} because of the readily available access point (AP) infrastructure. Regardless of the specific wireless communication technology, when multiple sensors are deployed in the human body, a WBAN in a star topology is usually created so that all the on-body nodes communicate with a gateway for forwarding the collected data. Fig.~\ref{fig:dual-rf-bcc-principle} (left) depicts the organization of a WBAN in a star topology that is employed by practical systems~\cite{survey-wban11}. Although RF is the only practical mechanism to forward data in this scenario, still several significant problems remain. RF signals suffer considerably from body shadowing in a highly variable way with respect to human body~\cite{ryckaert04,reusens07,nechayev05}. This makes communication between on-body nodes, and also off-body, very unreliable.

This inherent unreliability of RF communication around the human body is a critical problem for several real-life applications. We discuss two emerging application scenarios here to motivate our system design. One example is vital sign monitoring where multiple nodes need to be deployed in several different places of the human body~\cite{Yuce2010116}. Traffic is usually flowing in the uplink direction, i.e. from the WBAN nodes to the gateway. Consider for example a human that has a node on the torso and one on the wrist for monitoring the heart rate. If this human is a patient lying in bed, then the RF link of the sensor on the torso might be completely blocked. However, it is possible that the node on the wrist can communicate perfectly with the gateway. Or it can happen the other way around as Fig.~\ref{fig:dual-rf-bcc-principle}(a) indicates. Since humans usually move frequently, any sensor can be potentially blocked and the data may not be communicated on-time to the gateway (or it can be lost completely).
Besides health monitoring, the other important application is real-time media entertainment. With the rising popularity of wearable devices,\footnote{E.g. the Google glass project: www.google.com/glass} it is even possible that high data rate video streams need to be forwarded from the human body (uplink). One specific application example with downlink traffic, is audio streaming from a WiFi access point (AP) to on-body earphones without wires~\cite{baldus2010method}. In these two examples RF connectivity between the AP and the earphones/glasses might be unreliable leading to problems in the audio/video transmission. Therefore, the optimal node on the body must be found and act as a relay in order to transmit the data to and from the WiFi AP. This node can be for example a smartphone.

To address the problems identified in the aforementioned scenarios, in this paper we propose a novel WBAN architecture, a network layer (NWK) protocol that exploits the proposed architecture, and an optimization framework supported by an analytical performance model of the system. In our architecture that is depicted in Fig.~\ref{fig:dual-rf-bcc-principle}(b), all the WBAN nodes on the human body are equipped with both RF and body-coupled communication (BCC) transceivers. When the RF link of a node is unreliable the node uses the BCC link for forwarding its data through the human body to a node with a better RF link. The rationale of this design choice is based on the extremely power-efficient BCC technology that can be used instead of RF communication. Prototype BCC transceivers use simple baseband communication and have a very small form factor (1mm$^2$ dye in~\cite{fazzi09}). Thus, the proposed WBAN architecture with two transceivers on board a WBAN node is an economically and technically viable option for real-life applications.

\textbf{Contributions.} The contributions of this paper are:
\begin{itemize}
\item[C1] First, we propose a new cooperative WBAN architecture that employs two communication technologies namely BCC and RF. This architecture is orchestrated by a novel NWK relay selection protocol that identifies the optimal relay for forwarding the data off the body. The proposed scheme is unlike any existing relay selection protocols since two different communication technologies are used.
\item[C2] Second, for each medium access control (MAC) protocol of the RF and BCC subnetworks we develop accurate expressions of the delay and energy consumption as a function of specific protocol parameters.
Our novel contributions here include first the performance modeling of a NWK that concurrently uses two wireless MAC/PHY technologies. Second, our model is a cross-technology model, i.e. it quantifies the impact of a specific parameter setting for one MAC on the second and on the complete system.
\item[C3] Third, we propose the use of the analytical performance models for further optimization of the duty cycling of the BCC transceivers and the retransmission strategy of the RF transceivers so that energy consumption and delay are minimized.
\end{itemize}

\textbf{Benefits.} The \textit{first} advantage is that devices with a blocked RF link can use the services of the remaining nodes for forwarding their data to a gateway and improve thus the reliability of the data delivery. This idea is demonstrated with the help of Fig.~\ref{fig:dual-rf-bcc-principle}(b) while we have also outlined this generic architecture in~\cite{zubair}. This is accomplished with C1. \emph{Second}, RF transmissions are reduced to the absolute minimum since WBAN nodes do not communicate with each other through the RF link. The benefit is that the RF link is carrying a reduced load and the interference to surrounding RF devices is also minimized~\cite{baldus09,schenk08}. This is accomplished with C1. The \textit{third} advantage is that a node with an RF link that is not completely blocked, but operates inefficiently (higher number of required retransmissions that increase delay and energy consumption), can communicate with other nodes through BCC and select another node for data forwarding. This is accomplished with C2 and C3.
Our performance evaluation focuses on highlighting these benefits. Performance is evaluated both through simulations and also through the use of real RF traces. Results with IEEE 802.15.4 RF traces interestingly support even more the need for our system in real scenarios due to the severity of human body shadowing.

\textbf{Paper Organization.} The rest of this paper is organized as follows. Related work is presented in Section~\ref{section:related-works}. In Section~\ref{section:system-architecture} we provide an overview of the system architecture and in Section~\ref{section:relay-selection} we delve into the details of the relay selection and packet forwarding algorithm. Section~\ref{section:modeling-approach} presents the motivation and an overview of our performance modeling approach. In Section~\ref{section:rf-radio-model}  we present the performance model of the RF network while in Section~\ref{section:bcc-model} we present the analytical model of the BCC network. The MAC protocol optimization that further exploits the previous analysis is described in Section~\ref{section:optimization}. The performance of our system is evaluated with simulations and real measurements in Section~\ref{section:experiments}. Finally, in Section~\ref{section:conclusions} we conclude this paper.

\section{Related Work}
\label{section:related-works}
In this paper we deal with the problem of optimizing data delivery from a WBAN to a gateway. The problem is challenging because communication between on-body and off-body nodes is very unreliable~\cite{shah06,ruzzelli07}. The most promising way to attack the problem is through cooperation between WBAN nodes~\cite{survey-wban11}. With cooperation the nodes that aid in forwarding the data of other nodes are the \emph{relays}. There is a plethora of works that focus on selecting the optimal relay in the general case of wireless sensor networks (WSN) by considering different optimization objectives, while fewer works have focused in WBANs.

\textit{The majority of research works investigated cooperation in WSNs/WBANs with the objective to minimize power consumption.} In~\cite{naveen11} the authors consider relay selection in WSNs for minimizing the power consumption of individual nodes. They evaluate the impact of the sensor wake up schedules on the consumed power, and then they propose incentive mechanisms for participating in the cooperative network. This is an optimization approach that perceives the nodes as independent competing entities that may help each other for relaying data to a gateway. A number of authors investigated the impact of cooperation on the power consumption and lifetime of the complete WBAN introducing thus a more holistic approach~\cite{ehyaie09,engel13}. In these two works the cooperating nodes communicate with each other through the RF links opportunistically when they cannot reach the gateway. However, a particular node may still not be able to reach another on-body node. To combat these intermittent problems, another parameter that was also investigated together with relay selection was power control. Power control effectively means topology control in the WBAN~\cite{elias12}, i.e. we can allocate more power to nodes that need it so that they can reach the rest of the network or the gateway. Even though this approach ensures optimal power allocation for given channel conditions, power may be wasted in nodes with poor RF links. In the category of research efforts that target power minimization we cannot identify solutions that eliminate the problem of a completely blocked RF link, but rather state-of-art solutions that allocate optimally the system resources to combat the RF inefficiencies. This is still unacceptable in scenarios where data from a specific node must be transmitted in real-time to the gateway.

Another significant number of research efforts focused on the problem of reliable communication and was motivated from experimental results. In~\cite{braem07} the authors first focus on obtaining an experimental characterization of the channel. From the obtained results, the use of statically assigned relay nodes in specific body locations was proposed. Nevertheless, specific on-body nodes may still fail or be occluded due to body shadowing. Another aspect that affects the WBAN performance is the mobility of the user. Mobility models for WBANs driven from experiments were studied in~\cite{nabi11b}. In this work the authors study the problem of user mobility and posture changes and how they affect WBAN performance. The solution is a multi-hop protocol that is aided by a single relay.

In contrast to all the previous works that look into each specific node as a potential relay, there is an option to use multiple WBAN nodes for improving the reliability and/or energy/power. This class of protocols, that also targets generalized wireless cooperative networks, proposes the creation of \emph{clusters} across groups of nodes in order to improve the diversity gain in the RF link by simultaneously transmitting the same information over different wireless paths~\cite{sadek05}. Nevertheless, there is the major problem of node synchronization in this case.
In cooperative schemes that involve many nodes we can also add recent works that apply advanced techniques like network coding~\cite{arrobo11,zhang12}.

Despite the architectural and protocol differences, all the previously described schemes share one common characteristic which is the use of a single RF communication technology. To alleviate the problems of a single technology two different ones can be used. The number of works in this area is quite limited. A communication protocol that uses two technologies was presented in~\cite{tlc}. In that work the authors proposed the use of two different RF bands namely the 433MHz and 2.4GHz. The 433MHz band is used for data aggregation whereas the 2.4GHz band is used for data forwarding to the gateway. Since the range of the 433MHz band is limited to approximately 2 meters around the node it is possible to improve the reliability and energy consumption by reducing the number of nodes competing for the same channel. Still, both links use RF for on-body communications and two independent RF bands are required. \emph{Thus, all the techniques we discussed in the last three paragraphs cannot address completely the RF reliability problem because of the fundamental problems of the wireless channel.}

Finally, we should mention that none of the previous works employed an in-depth analysis of the impact of the MAC protocol parameters on energy/delay/reliability of the complete cooperative network. We could only trace a limited number of works for non-cooperative systems that use analytical protocol performance models for further optimization. These approaches use the MAC model of IEEE 802.15.4 for power minimization at an individual node~\cite{secon09-802154-mac-modelling-opt,ma04}.

When compared to all the related works, our approach is significantly different. It focuses on using simultaneously the most reliable and most efficient RF link (in terms of delay/energy) for data forwarding since these two aspects are intertwined. Our solution consists of a new WBAN architecture and a novel cooperative protocol design that is supported by an analytical performance model. With our system even when at least one node has an active RF link the data from every WBAN node will reach the gateway. To the best of our knowledge no other WBAN system can guarantee that.

\section{System Architecture}
\label{section:system-architecture}
We consider a WBAN that consists of $N$ nodes where each node is equipped with an RF transceiver and a BCC transceiver. These two hardware components use different MAC/PHY protocols and are both controlled from a specialized NWK protocol. The nodes are organized into a single-hop BCC network, and a subset $N_r$ of them form a single-hop RF network as shown in Fig.~\ref{fig:dual-rf-bcc-principle}. All the nodes that are part of the RF network can connect to a gateway for forwarding the collected data. Each wireless link has different delay/energy characteristics due to channel variations. While only $N_r$ nodes are part of the RF network, all the $N$ nodes participate in the BCC network. Non-relay nodes transmit their packets through the BCC link to one of the relays, while the selected relay transmits wirelessly to the gateway the locally generated and forwarded data packets. Nodes that do not have a relay responsibility are allowed to put into sleep mode the RF transceiver for reducing the energy consumption. However, since it is possible that in the future they assume the role of an RF relay, they wake up periodically, they transmit an RF packet, and after they receive a response they update the received signal strength indication (RSSI) value of their RF link.

\begin{figure}[t]
\begin{center}
  \includegraphics[keepaspectratio,width = 1.0\linewidth]{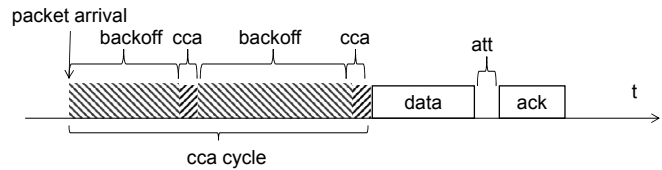}
  \caption{Packet transmission with the IEEE 802.15.4 MAC protocol.}
  \label{fig:rf-mac-protocol}
\end{center}
\end{figure}

\subsection{The IEEE 802.15.4 RF Network}
The IEEE 802.15.4 is used for RF communication. A non-beacon-enabled and un-slotted carrier sense multiple access with collision avoidance (CSMA/CA) algorithm is assumed for channel access. All the nodes sense the channel status during the clear channel assessment (CCA) slots. The basic idea of the un-slotted CSMA/CA algorithm is that backoff and packet transmissions are not aligned to specific slot boundaries. The un-slotted CSMA/CA operates as follows. Each time a device generates a packet for transmission it waits for a random number of slots, called the backoff counter, ranging from 0 to $2^{BE}-1$, where $BE$ denotes the backoff exponent. In IEEE 802.15.4, the backoff counter is decremented to zero regardless of the CCA result and by default $BE$ is initialized to 3. When the backoff counter reaches zero, the device performs CCA only once in order to check whether the channel is busy or not. If the channel is idle during the CCA period that has duration $T_{cca}$ (Fig.~\ref{fig:rf-mac-protocol}), the device transmits its data packet. When the channel is busy during the period $T_{cca}$, the backoff exponent $BE$ is increased by 1 and the random backoff procedure is repeated. The backslash-shaped blocks in Fig.~\ref{fig:rf-mac-protocol} depict the increase in the backoff time after a failed CCA. More details regarding its operation are provided later while the complete algorithm is available in~\cite{IEEE-802154}.

\begin{figure}[t]
\begin{center}
\includegraphics[keepaspectratio,width = 1.0\linewidth]{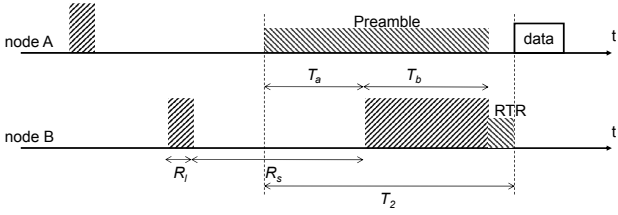}
  \caption{Low power listening and transmission of a preamble when the BCC channel is free. The time period $T_b$ indicates that the BCC node is awake and checks the channel for the presence of preambles.}
  \label{fig:bcc-mac-protocol2}
\end{center}
\end{figure}

\subsection{The BCC PHY and MAC}
BCC without requiring skin contact can be realized with two electrode transmitter/receiver devices capacitively coupled to the human body~\cite{baldus09}. The transmitter generates a variable electric field while the receiver senses the variable potential of the body with respect to the environment. A signal attenuation of less than 70dB has been measured between devices placed at various positions of the human body~\cite{schenk08}. The human body channel is especially affected by interference below 1MHz while for higher frequencies the interference level is below -75dBm. In this paper a BPSK modulation scheme is assumed where the digital pulses are directly transmitted to the human body through the capacitive plates~\cite{fazzi09}. Digital pulses of 1.2V for transmission are used, and an receiving band of 1-30MHz was chosen for improving the BER in the presence of interference. The BER was measured to be lower than $10^{-6}$ for the aforementioned conditions. The energy efficiency of the transceiver was measured at 0.32nJ/bit. Since the design of the PHY is not the focus of this paper we do not delve into further details. The interested reader can find more details about a typical prototype transceiver in~\cite{fazzi09}.

Regarding the MAC layer it is a protocol that employs the well-known concept of low power listening (LPL)~\cite{buettner06}. LPL exploits a specialized \textit{wake up receiver} hardware that is always in active mode and consumes very little energy while the main receiver is deactivated. Fig.~\ref{fig:bcc-mac-protocol2} presents the channel access scheme for a single packet. A preamble packet is always transmitted before the transmission of an actual data packet which means that nodes contend with preambles. The contention for the transmission of preambles follows the same backoff procedure with the RF MAC. Nodes that are asleep can wake up with the preamble and prepare for the actual data reception. The preambles contain the destination address so that they wake up the appropriate node. When a preamble is successfully received, the node that wakes up sends an acknowledgment packet that is named \emph{ready-to-receive} (RTR) since it also serves as an indication that the receiver can accept the actual data packet. After this transmission the receiver waits for the data packet to arrive and transmits the final ACK.

\begin{table}[!t]
\caption{Main functions used in the protocol stack of the complete system. Notation $tx\_X\_Y$ means that this function is executed when a message of type $X$ is transmitted/received in layer/technology $Y$. }
\label{tab1}
\begin{tabular}{|l|l|}
\hline
Main functions & Message contents\\
\hline
tx\_STATUS\_BCC(src~s) & RF/BCC param. for node $s$\\
tx\_TOKEN\_BCC(src~s,dst~d) &  Release token from node $s$\\
tx\_DATA\_BCC(src~s,dst~d,data~k) & TX data in the BCC link \\
tx\_DATA\_RF(src~s,dst~d,data~k) & TX data in the RF link \\
tx\_DATA\_NWK(src~s,dst~d,data~k) & Data from APPL to NWK \\
rx\_BCC(src~s) & Message received at the BCC\\
rx\_STATUS\_BCC(src~s) & RF stats received at the BCC\\
rx\_TOKEN\_BCC(src~s) & Token received at the BCC\\
rx\_DATA\_BCC(src~s) & Data received at the BCC\\
\hline
\end{tabular}
\end{table}

\section{Relay Selection}
\label{section:relay-selection}
\begin{figure}[!htb]
\framebox[3.6in]{
\begin{minipage}[t]{3.6in}
param\_est\_node($T_{est}$~sec)
\begin{algorithmic}[1]
\STATE Read latest local RF and BCC RSSI measurements
\STATE Estimate $\mathbb{E}[D_{rf}(i)],\mathbb{E}[D_{bcc}],\mathbb{E}[E_{rf}(i)],\mathbb{E}[E_{bcc}]$
\STATE tx\_STATUS\_BCC($\mathbb{E}[D_{rf}(i)],\mathbb{E}[E_{rf}(i)]^2/E_{rem}(i)$)
\STATE rx\_STATUS\_BCC()
\end{algorithmic}
------------------------------------------------------------------------------\\
rx\_STATUS\_BCC()
\begin{algorithmic}[1]
\STATE Update linked lists $NodesD,NodesE$
\STATE Sort ascending $NodesD,NodesE$
\STATE $better=0$
\FOR{relay $n \in Nodes$}
\IF{$  \mathbb{E}[D_{rf}(i)]> NodesD(n) $}
\STATE $better++$
\ENDIF
\ENDFOR
\IF{$  better \ge N_r \&\& I\_am\_relay==TRUE $}
\STATE // Drop relay status
\STATE $new\_relay \Leftarrow$ best non relay node from $NodesD$
\STATE tx\_TOKEN\_BCC(i,new\_relay)
\ENDIF
\IF{$  better < N_r \&\& I\_am\_relay==FALSE $}
\STATE // I should be a relay, waiting for TOKEN
\ENDIF
\IF{$  better \ge N_r \&\& I\_am\_relay==FALSE $}
\STATE power\_down\_BCC()
\ENDIF
\end{algorithmic}
------------------------------------------------------------------------------\\
rx\_TOKEN\_BCC()
\begin{algorithmic}[1]
\IF{$waiting\_for\_token==TRUE$}
\STATE //I am the new relay, I have the TOKEN
\ENDIF
\STATE //Update list $Relays$
\end{algorithmic}
------------------------------------------------------------------------------\\
tx\_DATA\_NWK(data $k$)
\begin{algorithmic}[1]
\FOR{relay $n \in Relays$}
\IF{$\mathbb{E}[E_{rf}(i)] > \mathbb{E}[E_{bcc}]+ NodesE(n)$ }%
\STATE //Forward pkt $k$ to the best relay node $n$
\STATE tx\_DATA\_BCC(i,n,k);  exit;
\ELSE
\STATE tx\_DATA\_RF(i,gway,k); exit; //Direct RF TX
\ENDIF
\ENDFOR

\end{algorithmic}
\end{minipage}
}
\caption{Pseudo-code for the cooperative protocol at the WBAN node $i$. }%
\label{fig:coop-protocol}
\end{figure}

The first and main task of our distributed WBAN system is to calculate the optimal subset of $N_r$ RF relays from all the available nodes, while the second task is to forward the data through the optimal relay if needed. To accomplish the first task, our system is designed as follows. Each WBAN node $i$ calculates the average delay and energy of the direct RF transmission and then it informs the remaining nodes in the network. The random variables of the delay and energy are denoted as $D_{rf}(i)$  and $E_{rf}(i)$ respectively. This functionality will be analyzed in subsections~\ref{subsection:1} and~\ref{subsection:2}. How this information is used for selecting a node as a relay will be examined in subsection~\ref{subsection:3}. The process of making the actual forwarding decision is described in~\ref{subsection:4}. All this functionality is concisely captured in the pseudo-code of Fig.~\ref{fig:coop-protocol}.

In the pseudo-code of Fig.~\ref{fig:coop-protocol} we introduce two specific \textit{control} messages at the NWK, namely $STATUS\_BCC$ for broadcasting the parameter estimates, and second $TOKEN\_BCC$ that is used for relay selection. A summary of the most important functions that are used by our protocol stack are provided in Table~\ref{tab1}. Finally, our protocol defines three data structures at each node in the form of linked lists, and are denoted as $NodesD,NodesE,Relays$ while their use will be explained in the next subsections.

\subsection{Parameter Estimation and Information Exchange}
\label{subsection:1}
The parameter estimation and information exchange functionality is captured in the function param\_est\_node() in the pseudo-code of Fig.~\ref{fig:coop-protocol} and is executed periodically every $T_{est}$ seconds by every node. When an RF node is not a relay, the RF transceiver is deactivated by entering an idle state~\cite{cc2420} and so it has to enter an active state to execute this function. The channel estimation itself is a local task at each RF node that is executed by default in IEEE 802.15.4 for every received packet and so no extra overhead is introduced by our scheme (i.e. when param\_est\_node() is invoked it reads the latest RSSI measurement from the related register of the micro-controller). Next, this measurement is used for calculating the average delay and energy that are denoted as $\mathbb{E}[D_{rf}(i)]$  and $\mathbb{E}[E_{rf}(i)]$ respectively (the expectation of these random variables). This is accomplished by using the RSSI measurement in the analytical model that we develop in Sections~\ref{section:rf-radio-model} and~\ref{section:bcc-model}. Next, node $i$ broadcasts the RF link performance estimates expressed though the average delay $\mathbb{E}[D_{rf}(i)]$ and average energy cost $ \mathbb{E}[E_{rf}(i)]^2/ E_{rem}(i)$,\footnote{For the energy cost we use the ratio $ \mathbb{E}[E_{rf}(i)]^2/ E_{rem}(i)$. This is a simple metric but it captures the fact that the cost is increased as the remaining energy is reduced after continuous utilization, while if the energy of a single transmission is low then this has an impact on lowering the cost. Our framework can support different and potentially better cost functions.} with the special message $STATUS\_BCC$. This message is transmitted at the BCC link and is broadcast which means that all the BCC transceivers of all the nodes will receive it. This concludes the description of the core functionality of the parameter estimation and information exchange functionality of our system. Next we discuss some important details.

\textbf{Robustness of STATUS\_BCC transmissions.}
The BCC PHY offers a proven reliable link since it transmits with BPSK modulation and in addition it uses channel coding for eliminating bit errors~\cite{fazzi09}. Furthermore, $STATUS\_BCC$ messages are sent with a constant period $T_{est}$ even when the channel does not change. In case of a packet loss, this periodic propagation of information allows all nodes to reach the same consistent state.

\textbf{Communication Overhead.} The forwarding of this information through the BCC link consumes a fraction of the BCC data rate. Here we provide a simple analysis to demonstrate that this communication overhead is negligible for the BCC link. Every $T_{est}$ the $N$ nodes transmit a $STATUS\_BCC$ message of length $L$ bits. And in the (unrealistic) worst case every $T_{est}$ all the current $N_r$ relays will also send a message $TOKEN\_BCC$ in order to change their role. Then the BCC data rate overhead is $(N +N_r)L/T_{est}$ bps. For a typical update frequency of $T_{est}$=1 sec, and a payload of $L$=20 bytes for these short control messages,\footnote{IEEE 802.15.4 with 16-bit addressing uses a 13-byte header.} $N$=$10$, and $N_r$=$5$, we have an overhead 2.4Kbps which is negligible for the BCC PHY of 10Mbps. The advantage of our system is that the BCC link offers highly reliable, low-energy, and high-data rate communication~\cite{baldus09,schenk08}. Therefore, the control messages that are exchanged across the different nodes and pass through the BCC link are more efficient in terms of both energy/bit and delay/bit when compared to RF.

\textbf{Frequency of Channel Estimate Propagation.} When the RF channel changes rapidly, more frequent updates will be required since they will affect the relay selection. Instead, when the channel is fairly static, the relay nodes will change in larger time scales. This is an important aspect that is a general problem that presents itself in wireless communication systems. In this paper we evaluate this aspect in our performance evaluation section.

\subsection{Processing Information Updates}
\label{subsection:2}
While collecting the information and broadcasting it is an important task, it is also critical to ensure that the incoming information at a node is properly processed. This functionality is implemented in the rx\_STATUS\_BCC() function of the pseudo-code in Fig.~\ref{fig:coop-protocol}. This function is invoked when a new $STATUS\_BCC$ message is received. In this case this function first re-sorts the two linked lists named $NodesD$ and $NodesE$ that contain all the $N$ network nodes according to the delay and energy estimates of their RF links. Next we explain why this simple message processing approach is enough.

\begin{figure}[t]
\includegraphics[keepaspectratio,width = 1.0\linewidth]{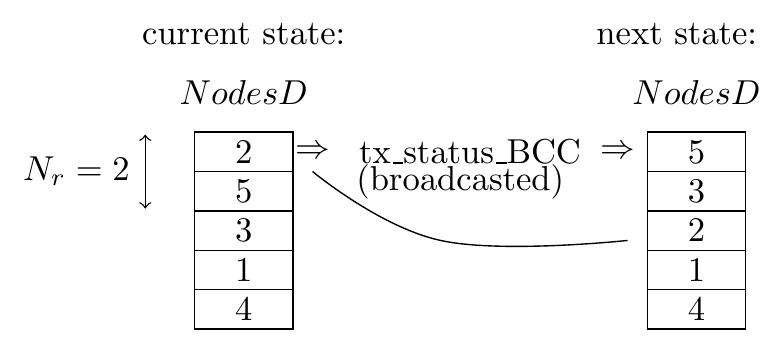}
\caption{Example that shows the ordered list $NodesD$ at all nodes. We assume that node 2 will stop being a relay. The transmission of $STATUS\_BCC$ that originates from node 2 will update $NodesD$ at every node.
}
\label{fig:example-status-bcc-message}
\end{figure}

\textbf{Exploiting the MAC for Serializing Information Updates.}
The $STATUS\_BCC$ messages are transmitted and processed sequentially by all nodes and this ensures consistent and reliable flow of information in the network.  Fig.~\ref{fig:example-status-bcc-message} has an example with five nodes that will clarify how these messages are processed. Assume that the nodes start from the same consistently order lists $NodesD$ and $NodesE$ shown in the left of Fig.~\ref{fig:example-status-bcc-message}. Assume now that node 2 measures a local change in the RF performance. From the last known state shown in the left of Fig.~\ref{fig:example-status-bcc-message}, it knows its new position in the list $NodesD$ that it now updates locally. Then, it broadcasts a $STATUS\_BCC$. The transmission of this message is naturally serialized by the BCC MAC in the sense that if there are more updates from other nodes they will be transmitted sequentially. Thus, this list will be updated only once for every $STATUS\_BCC$ message transmission and all nodes make this update simultaneously.

\subsection{Relay Selection Algorithm}
\label{subsection:3}
At this stage of our system description we have explained how all the necessary information is being updated. Now the actual relay selection decision is carried out in the second part of the function rx\_STATUS\_BCC(). In lines 4-8 of the pseudo-code node $i$ checks how many nodes perform better than itself in terms of delay and energy costs. A relay node drops the role of being a relay if it estimates from these measurements that either its delay or energy performance of the RF link is worse from more than $N_r$ (nodes that are allowed to be relays at all times). If this is the case it means that at least $N_r$ nodes have better performance characteristics depending on the delay or energy optimization objective. The node releases the token by transmitting a message $TOKEN\_BCC$ to the currently best non-relay node in list $NodesD$ (lines 11-12). Similarly, in the case that a node estimates that it should become a relay, but currently it is not a relay, it waits for the broadcasted token that will be released in the way we explained before (lines 14-16). The rx\_TOKEN\_BCC() function is responsible for processing the $TOKEN\_BCC$ messages. When this message is received by the appropriate node, it starts this new relay role.

\textbf{Robustness of TOKEN\_BCC transmissions.}
\label{subsection:4}
$TOKEN\_BCC$ messages are transmitted reliably, i.e. all nodes ACK the packet.\footnote{Recall that the transmission of any BCC packet is preceded by a preamble that wakes up the desired set of destination nodes. This preamble contains this ordered list of nodes, and after the data is transmitted they send sequentially ACKs. This is a mechanism of aggregating MAC protocol data units similar to IEEE 802.11n~\cite{IEEE-80211n}.} Since a token is acknowledged by the new relay, the "old" relay that releases it, and the remaining network nodes, know that the transaction was executed. Thus, all nodes know exactly which node is a relay and which is not. Note also that the transmission of tokens is also serialized in the BCC network which means that a new token is released when the last transaction is successfully completed.

\subsection{Packet Transmission and Forwarding}
\label{subsection:4}
At this stage in describing our system we have explained how it creates a list of the best relays. Now the task of node $i$ is to check if packet forwarding through relay node $n$ should take place and this task is executed at the network layer by the function tx\_DATA\_NWK(). When the application layer (APPL) wants to send a packet it does so with this function that is executed on demand. This is where the algorithm determines if the BCC or the RF links will be used. In the pseudo-code in Fig.~\ref{fig:coop-protocol}, the condition for forwarding data through a relay $n$ is that the expectation of the energy cost of the local RF transmission denoted as $\mathbb{E}[E_{rf}(i)]$, must be higher than the energy cost of the combined use of the BCC link and the RF link of the relay $n$:
\begin{equation}\label{eq:rf_bcc_energy_comparison}
\mathbb{E}[E_{rf}(i)] \geq \mathbb{E}[E_{bcc}]+ \mathbb{E}[E_{rf}(n)]^2/E_{rem}(n)
\end{equation}
The forwarding condition can also be formulated in terms of the delay and be expressed as
\begin{equation}\label{eq:rf_bcc_comparison}
\mathbb{E}[D_{rf}(i)] \geq \mathbb{E}[D_{bcc}]+\mathbb{E}[D_{rf}(n)],
\end{equation}
or the forwarding decision can be based on a combined metric (e.g forward only when both the delay and energy of the relayed transmission is better). In the previous expressions, $\mathbb{E}[E_{bcc}]$ and $\mathbb{E}[D_{bcc}]$ denote the expectation of the random variables of the energy and delay for one packet transmission in the BCC network.

Expressions \eqref{eq:rf_bcc_energy_comparison} and \eqref{eq:rf_bcc_comparison} are evaluated by the main loop of the packet forwarding algorithm depicted in function tx\_DATA\_NWK(k) of Fig.~\ref{fig:coop-protocol}. In this function, the loop checks first the best relay node that is contained in the list $Relays$, and if the condition is satisfied, then the message is forwarded to it. Otherwise, if the condition is not satisfied by the best relay then we employ direct RF transmission since all the remaining relays will have worse energy transmission cost.

\section{Cross-Technology MAC Performance Modeling}
\label{section:modeling-approach}
\textbf{Motivation.} In the previous section we explained that the RF relays are selected based on their delay and/or energy efficiency by using information that is distributed in the WBAN. If a node simply measures the average delay in the RF link and reports it to the WBAN, this does not account for the time consumed in retransmissions and channel access failures. The same is true for the BCC link. Thus, a node needs to precisely account for the previous parameters in order to decide if it will use forwarding through BCC or not. Therefore, before an actual forwarding decision is made by a node it must know that this is indeed the optimal course of action.
\textit{To accomplish that we develop not a cross-layer, but a same-layer and cross-technology performance model of the two MAC/PHYs that are controlled by the NWK.}

\textbf{Traffic Assumptions.} Throughout this section we consider applications where a node $n$ generates locally and asynchronously data packets with an Poission rate of $\lambda_n$ packets per second. Thus, we adopt an M/G/1 queueing model for modeling the random variables of the $D_{rf}$ and $D_{bcc}$ for the RF and BCC networks respectively. Several works that we reference in our subsequent analysis use a similar approach for capturing the service behavior of CSMA/CA algorithm. We also assume that the total load $\lambda= \sum^N_{n=1} \lambda_n $ can be forwarded even by a single RF node. The rate of packets that are locally generated and forwarded \textit{directly} from the relays is denoted as $\lambda_d=\sum_{n=1}^{N_r}\lambda_n$. We also denote with $\lambda_f$ the total network load\footnote{Note that although the total load that will be transmitted in the BCC network will be $\lambda_f$, it is not distributed equally to all the relay nodes since their link quality may be different instantaneously resulting into differences in~\eqref{eq:rf_bcc_energy_comparison} and~\eqref{eq:rf_bcc_comparison}.} that nodes receive through the BCC link and must \textit{forward} through their RF links, i.e. it is $\lambda_f=\sum_{n=N_r+1}^{N} \lambda_n$.  By decomposing the total load into the two components we have that $\lambda=\sum_{n=1}^{N_r}\lambda_n+ \sum_{n=N_r+1}^{N}\lambda_n$

\textbf{RF Channel Assumptions.} Similar to other works, that focus primarily on accurate modeling of IEEE 802.15.4, we adopt the following assumptions for modeling the radio network~\cite{kim2008:performance-802154}. The wireless channel is a frequency-flat Rayleigh fading wireless link that remains invariant per PHY frame, but may vary between frames. This means that the received signal-to-noise ratio (SNR) $\gamma$ per frame is a random variable. SNR estimation is performed from the RSSI at the receiver. Finally, we assume additive white Gaussian noise (AWGN).

\section{Performance of the RF Network}
\label{section:rf-radio-model}
Modeling the delay and energy cost of wireless RF transmissions is important for making the optimal decisions. However, existing models cannot be directly used~\cite{kim2008:performance-802154,bianchi-80211,wu02,Miluzzo08radiocharacterization}. The reason is that we consider the aggregate traffic load that is flowing from the WBAN to the gateway and at the same time this traffic goes though the different packet erasure links of the involved RF relay nodes. Thus, the basic idea of our model is that the packet loss probability is calculated by considering first the different packet erasure probability of each RF link, and second by using the aggregate traffic load for deriving the probability of having a CCA failure.

\subsection{Packet Erasures}
\label{section:erasures}
Now according to the IEEE 802.15.4 PHY coding scheme, the code that is used is a pseudo-orthogonal scheme where 4 bits are encoded together into a 32 chip signal.
The modulation scheme is offset quadrature phase shift keying with half sine shaping (OQPSK-HSS) at a rate of 2Mchip/s. The result is that the 1/8th rate coding scheme achieves a throughput of 250kb/s. Now if $Q(x)$ is the Gaussian $Q$ function, then for OQPSK in a Rayleigh channel, the instantaneous bit error rate (BER) is:
\begin{equation}\label{eqn:1}
    \epsilon= Q \Big ( \sqrt{3\gamma}  \Big )
\end{equation}
The packet error probability can be calculated as
\begin{equation}\label{eqn:3}
peb=1-(1-\epsilon)^{L},
\end{equation}
where $L$ is the length of the packet in bits. Since ACKs are used the probability that a PHY frame transmission fails towards the gateway from relay node $n$ is:
\begin{eqnarray}\label{eqn:2}
\pi_{e}(n) = 1- [1-peb_{data}][1-peb_{ack}],%
\end{eqnarray}
where $peb_{data}$ and $peb_{ack}$ are the packet error probability for data packets and acknowledgments respectively.

\subsection{CCA Failures}
Let now $\pi_{cca}$ denote the probability that the RF channel is busy when CCA is performed. The probability that the channel is free after $u-1$ failed CCA attempts is simply $\pi_{cca}^{\upsilon-1}(1-\pi_{cca})$. The value of $\pi_{cca}$ depends only on $N_r$, i.e. how many nodes contend for the RF channel. This probability is independent at each attempt while such an approximation is very accurate for saturated traffic as shown in~\cite{bianchi-80211}, and has been widely adopted in the literature also for unsaturated traffic~\cite{wu02}. We provide the methodology for the derivation of $\pi_{cca}$ below.

Let $S_{rf}$ denote the number of served/transmitted packets in a busy period of the M/G/1 RF queueing system. Then it is
\begin{equation}\label{rho_rf}
\rho=\frac{\lambda}{\mu_{rf}}, \quad \mathbb{E}[S_{rf}]=\frac{1}{1-\rho}.
\end{equation}
Note that for the M/G/1 queuing system we consider the total load that is transmitted in the RF network with an average service rate $\mu_{rf}$. This is because there is no need to differentiate the traffic from each node since the aggregate is what matters for the model as along as the arrival rate at each node is Poisson. Now for each packet it will be that the average service rate is:
\begin{equation}\label{mu_rf}
\mu_{rf}=1/(\mathbb{E}[D_{HOL}]+ T_{d} +T_{att} + T_{ack})
\end{equation}
In the above expression $D_{HOL}$ is the random variable of head of line (HOL) delay while $T_d$ and $T_{ack}$ are the time durations of the data packets and the ACKs. $T_{att}$ is the required turnaround time to switch from receiving (RX) to transmitting (TX) mode as specified in the standard~\cite{IEEE-802154}. The HOL delay is the duration from the time instant that the packet arrives at the head of the RF transmission queue to the time instant just before its transmission or final discard.

To obtain the expression for $\pi_{cca}$ we have to divide the time period that the channel is occupied by the remaining $N_r-1$ devices versus the total duration of the M/G/1 system busy period (see Fig.~\ref{fig:rf-mac-protocol}). The term $\frac{1}{\lambda_f+\lambda_d}+\mathbb{E}[S_{rf}]\mathbb{E}[D_{HOL}]$ is the total duration of the busy period in the RF network. So we have that:
\begin{equation}\label{pi_cca}
\pi_{cca}=\frac{(N_r-1)(1-\pi_{loss})\mathbb{E}[S_{rf}](T_{cca}+ T_{d} + T_{ATT} + T_{ack})}{\frac{1}{\lambda_f+\lambda_d}+\mathbb{E}[S_{rf}]\mathbb{E}[D_{HOL}]}
\end{equation}
In the above equation we can replace $\mathbb{E}[S_{rf}]$ that can be derived as a function of $\mathbb{E}[D_{HOL}]$ from~\eqref{rho_rf} and~\eqref{mu_rf}. In the last equation, $\pi_{loss}$ is the average packet loss probability in the RF network that occurs both because of erasures and CCA failures. Since these two events are independent, we have that
\begin{equation}\label{P_loss}
    \pi_{loss}=\frac{1}{N_r}\sum_{n=1}^{N_r} \sum_{k=0}^{M_r-1}\pi_{e}^{k}(n)(1-\pi_{e}(n)) \Big (1- \sum_{\upsilon=0}^{M_c-1}\pi_{cca}^{v}(1-\pi_{cca}) \Big ),
\end{equation}
where $M_r$ and $M_c$ are the maximum number of transmissions and CCA attempts respectively. Note in the last equation that we account for the different $\pi_{e}(n)$ that an RF relay might have. \textit{This is an important difference with related works on MAC performance modeling~\cite{kim2008:performance-802154,bianchi-80211,wu02,Miluzzo08radiocharacterization} that consider a single packet erasure probability but instead we consider the different impact of all the used relays.}

Now we will also derive a second equation that along with~\eqref{pi_cca} can be solved numerically for the derivation of $\pi_{cca}$ and $\mathbb{E}[D_{HOL}]$.

\subsection{Delay}
For calculating the delay $D_{HOL}$ we have to model the behavior of the backoff algorithm. We denote the value of the contention window as $W_i=2^{BE_{min}+i}$. For the successful CCA after $M_c-1$ failed ones, the HOL delay that accounts only for CCAs will be:
\begin{eqnarray}\label{D_cca}
\mathbb{E}[D^{cca}_{HOL,rf}] = \sum_{\upsilon=0}^{M_c-1}\pi_{cca}^{v}(1-\pi_{cca}) \Big \{ \sum_{i=0}^{\upsilon}\frac{W_i-1}{2} T_{s}\nonumber\\
    +(\upsilon+1) T_{cca} \Big \}
    +\pi^{M_c}_{cca} \Big \{ \sum_{i=0}^{M_c-1}\frac{W_i-1}{2} T_{s}+(M_c+1) T_{cca} \Big \}
\end{eqnarray}
The previous equation covers the case of $v$ failed CCA attempts and a last successful one.  In the case that there were $v$ times a failed CCA, the total delay is attributed to the backoff algorithm that was executed at the sender, plus there is the time spent for the $v$ CCAs and is denoted as $T_{cca}$ (plus one successful CCA). The parameter $T_{s}$ is the duration of the slot time. The second summation term covers the case of reaching the maximum limit of $M_c$ failed CCAs.

If there is an erasure and the number of allowed transmissions attempts is $M_r$, then the average HOL delay becomes:
\begin{eqnarray}\label{D_e}
\mathbb{E}[D_{HOL,rf}]&=&\frac{1}{N_r}\sum_{n=1}^{N_r}\sum_{k=0}^{M_r-1}\pi_{e}^{k}(n)(1-\pi_{e}(n))  \{ \mathbb{E}[D^{cca}_{HOL,rf}] \nonumber\\
&+&k (T_{att}+T_{d}+T_{ack})  \}
\end{eqnarray}
The term $\pi_{e}^{k}(n)(1-\pi_{e}(n))$ is similarly with before the probability that a packet was erased $k$ times before a final successful transmission. If the data packet is lost then the lack of an ACK incurs a delay $T_{att}+T_{ack}$. The transmitter observes the lack of an ACK packet and assumes the packet was lost and proceeds with backoff and another round of possible multiple CCAs (depicted in Fig.~\ref{fig:rf-mac-protocol}). Now the average per packet delay of the RF link is:
\begin{equation}\label{D}
    \mathbb{E}[{D}_{rf}]=\mathbb{E}[D_{HOL,rf}]+ T_{d} +T_{att} + T_{ack}
\end{equation}

\subsection{Energy of RF Transmissions}
In this subsection we characterize the average energy consumption of the RF transmissions. If the power consumption of the RF subsystem in active, CCA, transmit, and receive modes are $P_{act},P_{cca},P_{tx}$, and $P_{rx}$ respectively, then the energy required for the RF transmission is calculated as follows:
\begin{eqnarray}\label{E_rf_tot}
    \mathbb{E}[E_{rf}]&=&(\mathbb{E}[D_{HOL,rf}]-A_{cca} T_{cca}) P_{act}+P_{cca} A_{cca} T_{cca}\nonumber\\
        &+&T_{d} P_{tx}+T_{att} P_{act}+T_{ack} P_{rx}
\end{eqnarray}
The first term is the energy consumed during contention while the node was simply in active mode, but we subtract the total time duration that CCAs were executed during the total time that the packet was in the HOL position. $A_{cca}$ is the average number of CCAs performed during this period and it is equal to
\begin{equation}
A_{cca}=\frac{1}{N_r}\sum_{n=1}^{N_r} \big( \sum_{k=0}^{M_r-1} k \pi_e^k(n) (1-\pi_e(n))\sum_{u=0}^{M_c-1} u \pi^u_{cca} (1-\pi_{cca}) \big ).
\end{equation}

\section{Performance of the BCC Network}
\label{section:bcc-model}
Although the transmission through the BCC network is very reliable, an increased number of contending nodes will eventually translate to an increase in the delay and energy consumption of the BCC subsystem. Therefore, in this section we analyze the MAC layer performance of the BCC network in order to calculate the impact of the different parameters on the packet delay and energy.

\subsection{Delay}
We distinguish the delay to transmit a packet successfully on the BCC network as three random components. From Fig.~\ref{fig:bcc-mac-protocol2} we can identify those three components easily. First, $\mathbb{E}[D_{HOL,bcc}]$ is the expectation of the random HOL delay spent by the transmitter node from the moment it decides to send a data packet, until the time instant it transmits a preamble packet. This delay does not include the processing time and the transmission time of the preamble. Then $T_2$ is the random delay spent by the transmitter node until the receiver node is in the listening state and an RTR packet arrives at the transmitter node. Finally, $T_3$ is the random delay spent by the transmitter node from the instant of the RTR reception until the transmission of the data packet. This component also includes the processing time, the transmission time of the data packet, and the transmission of the ACK.

\textbf{HOL Delay Before Preamble Transmission.} With the used MAC protocol, the preamble transmission mechanism is based on a binary exponential backoff algorithm similar to the IEEE 802.15.4 RF MAC. As we explained earlier, once the BCC channel is obtained through this preamble contention scheme, data packets are transmitted contention-free. If the channel is busy, a random backoff is executed before a further preamble transmission attempt. Let $M_{mp}$ be the maximum number of CCA attempts for a preamble. Also $T_{cca}$ is the required channel sensing time from the BCC hardware and $T_{s}$ is the slot time. Let also $\sigma_{cca}$ be the probability to sense the BCC channel busy, and $S_{bcc}$ is the number of transmitted packets in a busy period of the BCC queueing system. The parameter $\sigma_{cca}$ in this case is calculated in the same way with the RF network:
\begin{equation}\label{sigma_cca}
\sigma_{cca}=\frac{(N-N_r-1)(1-\pi_{loss})\mathbb{E}[S_{bcc}](T_{cca}+ T_{d} + T_{att} + T_{ack})}{\frac{1}{\sum_{n=N_r+1}^{N}\lambda_n}+\mathbb{E}[S_{bcc}]\mathbb{E}[D_{HOL,bcc}]}
\end{equation}			
Recall that in this case there are no packet erasures since reliable transmission is assumed once the channel is occupied and so $\pi_{loss}$ is calculated from a simpler version of~\eqref{P_loss}. \textit{Equation~\eqref{sigma_cca} demonstrates another novel aspect of our model since it incorporates this cross-technology interaction: It quantifies the impact of different number of RF relays $N_r$ on the probability that there is a channel access failure in the BCC link. This expression also quantifies the impact of a different forwarding load $\lambda_f$ on the BCC network.} Similar with the RF case we can write for the transmission of the preambles:
\begin{eqnarray}\label{eq:T1_avg}
   \mathbb{E}[D_{HOL,bcc}] &=&  \sum_{\upsilon=0}^{M_{mp}-1} \sigma_{cca}^{u}(1-\sigma_{cca}) \Big \{ \sum_{i=0}^{\upsilon}\frac{W_i-1}{2} T_{s}\nonumber \\
&+&     (\upsilon+1) T_{cca}\Big \}  +\sigma_{cca}^{M_{mp}}\Big \{ \sum_{i=0}^{M_{mp}-1}\frac{W_i-1}{2} T_{s}\nonumber\\
&+&(M_{mp}+1) T_{cca} \Big \}
\end{eqnarray}
Since the BCC MAC is also modeled as an M/G/1 queue, we can have a similar expression with~\eqref{pi_cca} for $\sigma_{cca}$ and $\mathbb{E}[D_{HOL,bcc}]$. This second equation is solved jointly with~\eqref{eq:T1_avg}, expressions for the HOL delay and channel busy probability $\sigma_{cca}$ for the BCC network are obtained.

\textbf{Preamble and RTR Transmission Delay $T_2$.} Now we calculate $T_2$, i.e. the random delay the transmitter node waits from the start of transmitting a preamble until the RTR transmitted by the receiving node is received by the transmitter node. Fig.~\ref{fig:bcc-mac-protocol2} depicts this important detail for the calculation of $T_2$. An important characteristic of the used BCC MAC protocol is that the selection of the sleeping time $R_s$ is always such that the target sleeping node can be awake only from the successful transmission of a single preamble packet. The listening time $R_l$ on the other hand depends on the hardware. Since the preamble has a fixed duration, the receiver will always wait until the end of the preamble~\cite{buettner06}. Therefore it will be:
\begin{equation}\label{T_2}
T_2=T_{pream}+T_{att}+T_{rtr}
\end{equation}
Although $T_2$ was derived with simple reasoning, there is a need to calculate another parameter that will be needed for the energy estimation in a later subsection. To this aim we define as $T_w$ the random time to wait until the wake up receiver and the actual receiver wakes up completely. In our system this time duration can be easily modeled since it falls with uniform distribution in the range $[0,R_s]$. This is because such a time duration is computed from the beginning of the preamble transmission of the transmitting node that may uniformly fall somewhere in the interval $[0,R_s]$. Therefore, the average value for $T_w$ will be $\mathbb{E}[T_w]$=$R_s/2$. We also denote with $T_a$ the random time to wait from the beginning of the preamble transmission until the start of the listening period (see Fig.~\ref{fig:bcc-mac-protocol2}). Now if the node is awake it is easy to see that $T_a$=$0$, otherwise it will be $\mathbb{E}[T_w]$ on average. So for the average value of $T_a$ we can write that:
\begin{equation}
\mathbb{E}[T_a]=\mathbb{E}[T_w] \Pr\{asleep\}=\frac{R_s}{2}\frac{R_s}{R_s+R_l}
\end{equation}
Also from Fig.~\ref{fig:bcc-mac-protocol2}, $T_b$ is the delay from the wake up moment of the receiver due to the preamble, until the end of the preamble. This time duration will be simply
\begin{equation}
\mathbb{E}[T_b]=T_{pream}-\mathbb{E}[T_a].
\end{equation}

\textbf{Data Transmission Delay $T_3$ and Total Delay.} After the node is awakened and the RTR is sent, the transmission of the data packet with the associated ACK consume a constant amount of time that depends only on the length of the data packet $L$. Since the probability of packet errors at the BCC PHY is assumed to be negligible the BCC transmission is always successful and $T_3$ is constant. This constant delay component is
\begin{equation}\label{T_3}
T_3=T_{d}+2T_{att}+T_{ack},
\end{equation}
where $T_{att}$ is the time required for the BCC transceiver to switch from RX to TX mode, and $T_d$ is the transmission time of the data packet of length $L$. Thus, we have that the total average delay from the moment a packet becomes the HOL packet until it is ACKed is:
\begin{equation}
\mathbb{E}[D_{bcc}]=\mathbb{E}[D_{HOL,bcc}]+T_{2}+T_{3}
\end{equation}

\subsection{Energy of BCC Transmissions}
The total energy consumption over a listening-sleeping time duration $R_s + R_l$, is calculated by adding the energy consumed by a transmitting node to send a data packet ($\mathbb{E}[E_{send,bcc}]$), and the energy consumed by the receiving node to receive the data packet ($\mathbb{E}[E_{recv,bcc}$). The calculations use the value of power consumption for the BCC transceiver and the time duration that this power drainage occurred, while we again remove the subscript that indicates the BCC system for avoiding clogging the presentation. From Fig.~\ref{fig:bcc-mac-protocol2} we derive the time instants that the sender uses its transceiver in RX, and TX modes.\footnote{Note that for the used BCC hardware it is $P_{cca}\simeq P_{rx}$~\cite{fazzi09}.} So it will be just for the contention round for a single packet transmission:
\begin{equation}\label{E_1_bcc}
    \mathbb{E}[E_{1,bcc}]= \mathbb{E}[D_{HOL,bcc}(act)] P_{act} +\mathbb{E}[D_{HOL,bcc}(cca)] P_{cca}
\end{equation}
Similarly we calculate $\mathbb{E}[E_{2,bcc}],\mathbb{E}[E_{3,bcc}]$ from~\eqref{T_2} and~\eqref{T_3}. The average energy consumption $\mathbb{E}[E_{send,bcc}]$ is the sum of these three expectations.

At the receiver we calculate similarly the time instants that the hardware is in TX/RX mode, active, and in sleep mode. During the reception of a preamble we calculated that the average duration that the receiver is awake is $\mathbb{E}[T_b]$. So we have that for a single packet transmission:
\begin{eqnarray}\label{E_recv_bcc}
    \mathbb{E}[E_{recv,bcc}] &=& R_s P_{sleep}+\mathbb{E}[T_b]  P_{act}+T_{rtr} P_{tx}\nonumber\\%
    &+&T_{d} P_{rx}+T_{ack} P_{tx}
\end{eqnarray}
The average BCC energy $\mathbb{E}[E_{bcc}]$ for a single packet transmission is the sum of~\eqref{E_recv_bcc} and $\mathbb{E}[E_{send,bcc}]$ that is derived as we explained a few lines above.

\section{Optimization}
\label{section:optimization}
From our derivations we observe from~\eqref{D_cca},~\eqref{D_e},~\eqref{E_1_bcc} and~\eqref{E_recv_bcc} that the total average energy consumption depends on several parameters. If we notice more carefully we see that for constant $N$ and $N_r$ the energy and delay expressions for the RF and BCC links are independent. This means that the energy for the RF link can be further optimized individually and locally by each node depending on the local RF channel while the parameters of the BCC link are optimized in the same way by all nodes. Many optimization objectives exist for such a system. As a representative case, we examine energy minimization subject to an average packet delay constraint.

\subsection{Energy Minimization}
For the RF part two parameters can be optimized for a given $N_r$ and packet length $L$ (this usually depends on the application), and these are the maximum number of transmissions $M_r$ and CCA attempts $M_c$. We use the RF average energy expression in~\eqref{E_rf_tot} for optimally selecting the two aforementioned parameters at relay node $n$:
\begin{eqnarray}\label{arg_min_E_rf}
    \min_{M_{c},M_{r}} \mathbb{E} [ E_{rf}](M_c,M_r,L,N_r,\pi_e(n))\\
    \text{subject to } \pi_{loss} (n) \leq 15\% \nonumber
\end{eqnarray}
The constraint in the above is the packet loss rate (PLR) for a specific node $n$ and is given from~\eqref{P_loss} before the averaging over all the relays. An important note is that in theory the above optimization problem must be solved on-line every time a node calculates a different $\pi_e$ with the help of~\eqref{eqn:1},~\eqref{eqn:3},~\eqref{eqn:2} and the local RSSI.

For the energy of the BCC link more parameters can be optimized while the optimization problem has \emph{to be solved once for a given nework configuration}. The reason is that for a given number of nodes $N$, relays $N_r$, and locally generated load by each node $\lambda_n$, the BCC load $\lambda_f$ that must be forwarded is constant regardless of who is the RF node that actually forwards it. This also means that $\sigma_{cca}$ is constant for a given network configuration. With this logic we can easily see that the following optimization problem indeed needs to be solved once:
\begin{eqnarray}\label{arg_min_E_bcc}
    \min_{R_l,R_s,M_{mp}} \mathbb{E}[E_{bcc}](R_s,R_l,M_{mp},L,\lambda_f,\sigma_{cca},N,N_r)\\
    \text{subject to } \mathbb{E}[D_{rf}]+\mathbb{E}[D_{bcc}] \leq \tau \nonumber
\end{eqnarray}
The constraint denotes the average estimate of the delay that must be less that the maximum allowed delay $\tau$. The reason that we need the delay constraint here unlike \eqref{arg_min_E_rf}, is that the energy in the case of the BCC link can be minimized by increasing $R_l,R_s$ which will incur higher delay.

\subsection{Solving the Optimization Problems Online}
Regarding the solution and the implementation, for the parameters $\pi_{cca},~\sigma_{cca}$, $D_{HOL,bcc}, D_{HOL,rf}$, we produce off-line several analytical expressions for a given $N_r$, $N$, and $L$. The result is that for the delay and energy of the RF network a linear expression is produced that only has as optimization parameters $M_c$, $M_r$, and accepts as input $\pi_e(n)$ from the real measurements of each node $n$. A linear equation with two unknowns can be solved in real-time even by underpowered micro-controllers. Also for the BCC network the energy and delay formulas are similarly calculated off-line for specific sets of input parameters, while $M_{mp}$,$R_s$,$R_l$, are the optimization parameters. After the selection of the optimal values is made, they are enforced in next cycle of the node.\footnote{A further optimization was employed in our actual implementation since $R_s,R_l$ are the same for the complete network. Only one node is needed in practice to make the previous optimization and it informs the remaining ones about the result.}

\begin{figure}[t]
\begin{center}
\subfigure[$\tau$=1s]{\includegraphics[keepaspectratio,width = 0.5\linewidth]{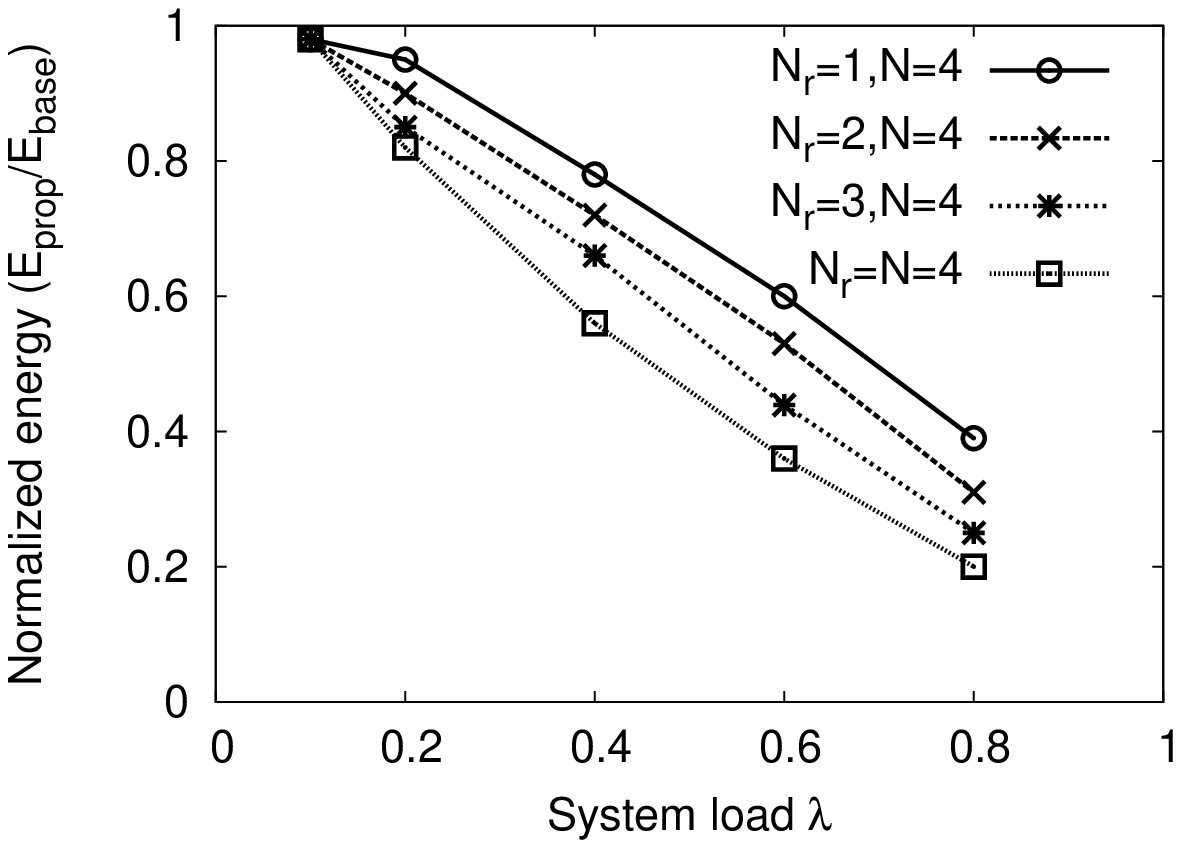}\hspace{-0.2cm}}%
\subfigure[$\tau$=1s, $N_r$=$N$=4]{\includegraphics[keepaspectratio,width = 0.5\linewidth]{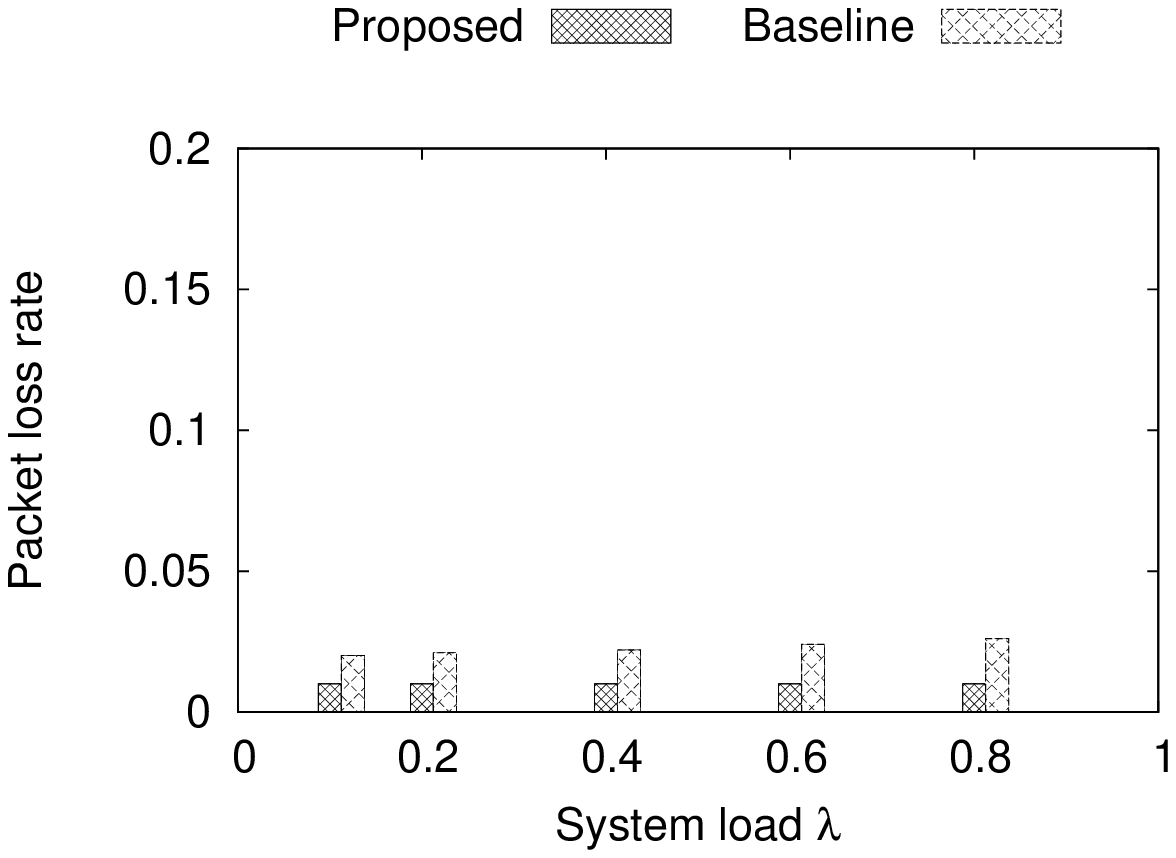}}
\subfigure[$\tau$=50ms]{\includegraphics[keepaspectratio,width = 0.5\linewidth]{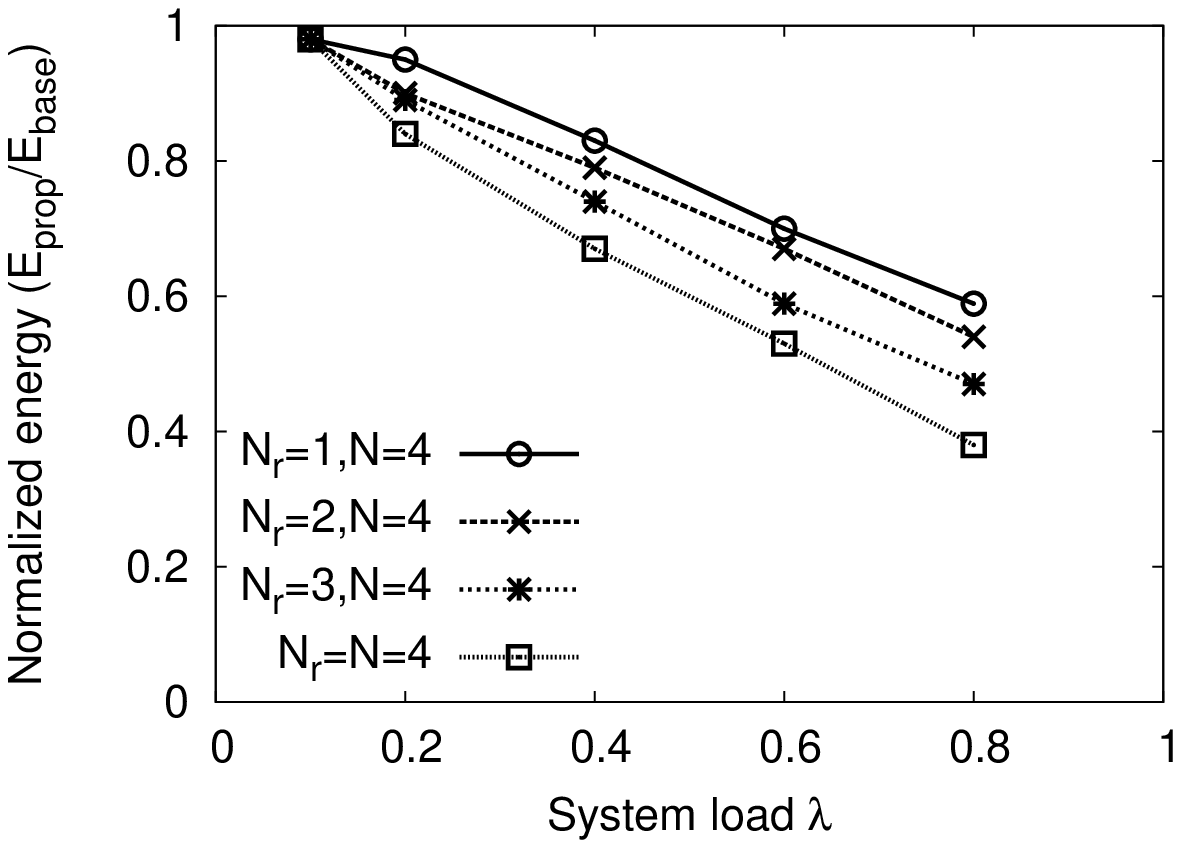}\hspace{-0.2cm}}%
\subfigure[$\tau$=50ms, $N_r$=$N$=4]{\includegraphics[keepaspectratio,width = 0.5\linewidth]{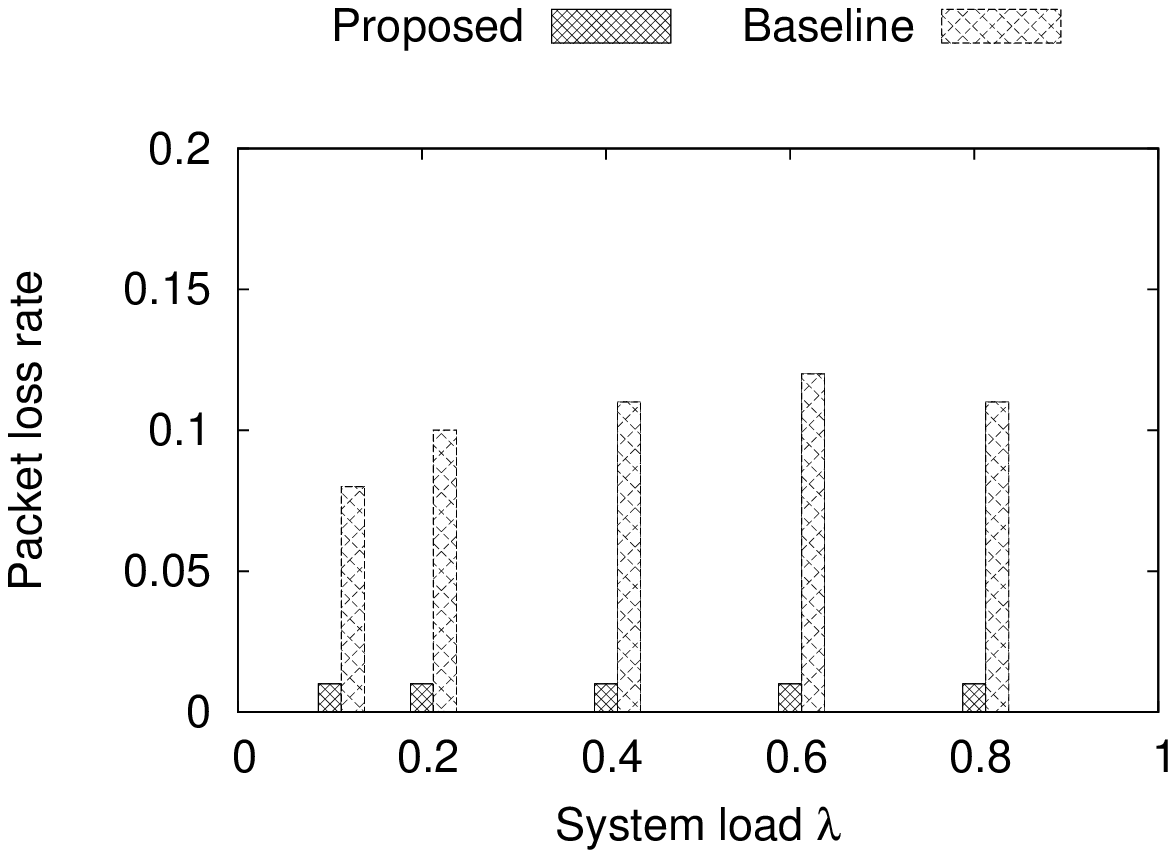}}
\caption{Results for the relative energy consumption and the PLR of the proposed system versus the baseline system for different $\lambda$ and \emph{Scenario 1}.}
  \label{fig:energy-vs-load}
\end{center}
\end{figure}

\section{Performance Evaluation}
\label{section:experiments}
\subsection{Simulation}
\textbf{Evaluated Systems.} In the first part of our evaluation of the proposed system we used simulation due to the lack of hardware devices equipped with RF and BCC transceivers that are in a size that can be deployed in a human. For simulating our system we implemented the NWK protocol and the RF and BCC MAC protocols. For the BCC PHY system we used parameters available from the literature and are presented shortly, while for the RF link we considered a Rayleigh channel and the IEEE 802.15.4 PHY that we described earlier in Section~\ref{section:system-architecture}. We compare our system with a baseline system that employs only RF communication. In this system when a node is completely blocked from the gateway, it connects opportunistically with an on-body node that can reach the gateway. For all systems we measured energy consumption, average per packet delay, and PLR.

\textbf{Parameters and Topology.} The raw data rate of the BCC transceiver is approximately 8.5Mbps with a BER of $10^{-6}$ and a voltage setting of 1.2V, the power consumption is 2.1mW in RX mode, and 0.6mW in TX mode~\cite{fazzi09}. For a typical IEEE 802.15.4 RF transceiver the CC2420 from Texas Instruments that operates with the same voltage, the current consumption is  at 19.7mA for the TX mode and 17.4mA for the RX mode~\cite{cc2420}. For this simulation every node starts with the same energy budget and there is a fixed number of 1000 packets that must be transmitted from each node to the gateway. We also set $L$=100 bytes, $T_{s,bcc}$=23us, $T_{ack,bcc}$=$T_{rtr,bcc}$=0.1ms, $T_{att,bcc}$=0.1ms, $T_{d,bcc}$=0.2ms, $T_{att,bcc}$=0.384ms, $T_{s,rf}$=0.192ms, $T_{cca,rf}$=0.25ms, $T_{d,rf}$=1.12ms, and $T_{ack,rf}$=0.352ms. Regarding the topology of the network, we consider \emph{Scenario 1} where a human sits in a static position with randomly placed nodes in the body but with more than 50\% of the nodes within line-of-sight (LOS) of the gateway. For \emph{Scenario 2} we considered a more challenging case where a human has half of the nodes in the torso (within LOS) and half in the back while lying in bed (no LOS).

\begin{figure}[t]
\begin{center}
\subfigure[Varying RF channel SNR for \emph{Scenario 1}]{\includegraphics[keepaspectratio,width = 0.5\linewidth]{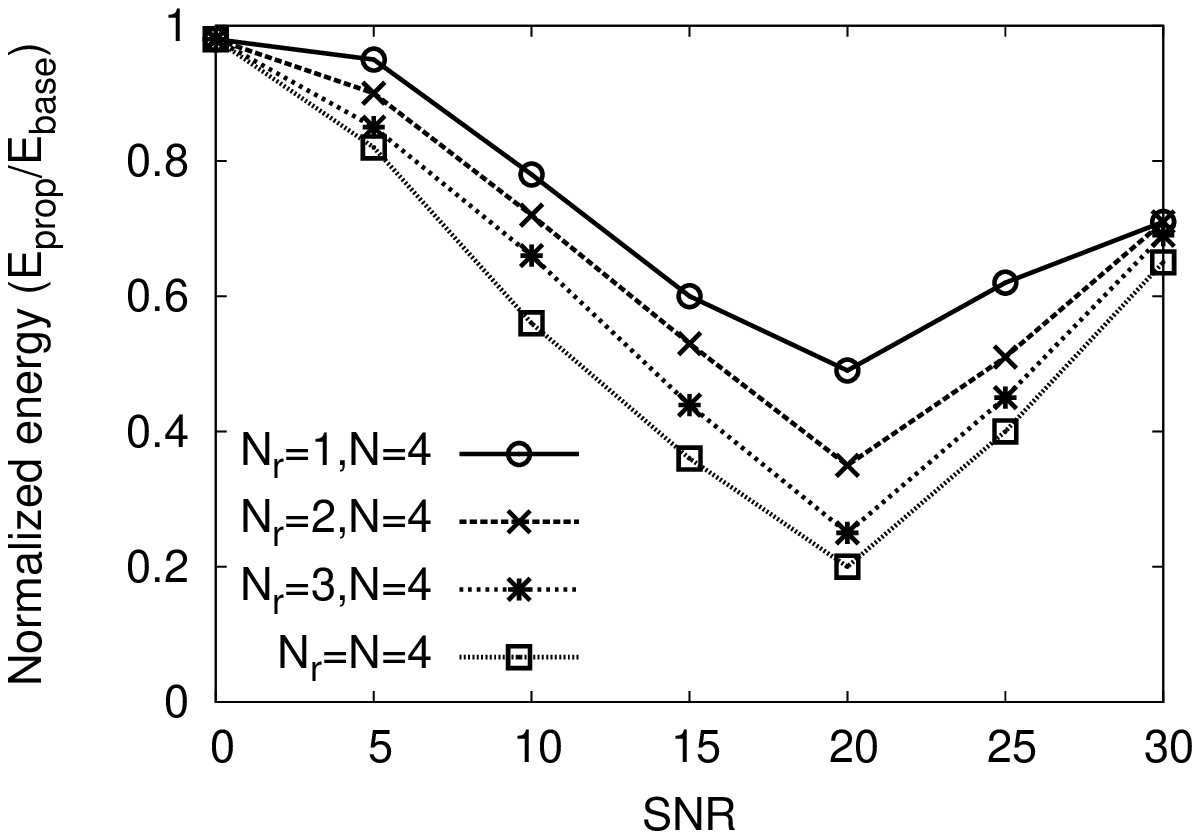}}%
\subfigure[$N_r$=2,$N$=4, varying $M_r$ and $M_c$, for \emph{Scenarios 1,2}]{\includegraphics[keepaspectratio,width = 0.5\linewidth]{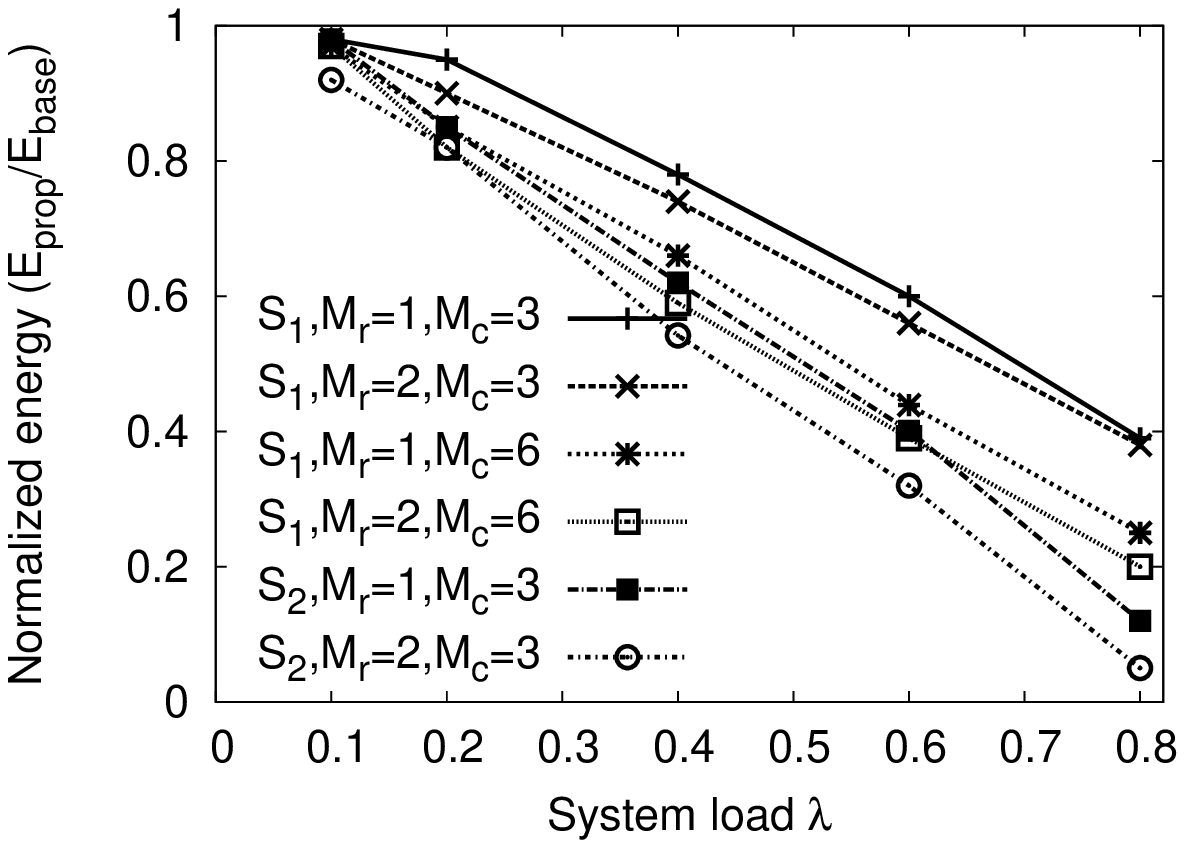}}
\caption{Results for the relative energy consumption of the proposed system versus the baseline system.}
\label{fig:energy-vs-snr}
\end{center}
\end{figure}

\subsubsection{Results for Energy Minimization}
Fig.~\ref{fig:energy-vs-load}(a) depicts the ratio of the consumed energy of the proposed system ($E_{prop}$), versus the energy of the baseline system ($E_{base}$), for a different system load $\lambda$ and \emph{Scenario 1}. The delay constraint is very relaxed and equal to $\tau$=1s while $M_r$=$M_c$=3. The maximum load corresponds to 250Kbps which is the maximum data rate at the PHY of the IEEE 802.15.4 RF link. It is interesting to see from these results that with the proposed system the energy reduction is significant regardless of how many relays ($N_r$) are used. The dominant factor that determines the reduction is the load which as it is increased, the performance gain becomes higher even with a single relay. This can be explained because for a higher load with the baseline method nodes with poor RF links are still used for transmission and successive failures (allowed by the high $\tau$) result in a higher number of retransmissions. The higher number of packets in the RF medium increases also the contention that has minor impact in the overall PLR as Fig.~\ref{fig:energy-vs-load}(b) shows. However, when the delay constraint is more tight, and equal to 5ms, the energy differences as we can see in Fig.~\ref{fig:energy-vs-load}(c) are reduced, but PLR increases dramatically as Fig.~\ref{fig:energy-vs-load}(d) indicates. For a high load, PLR becomes nearly 10\% for the baseline case while it is increased slightly in the proposed system. This is because the average number of retransmissions and CCA attempts are reduced significantly for the baseline case, which eventually leads to lower energy wastage but still high PLR. \emph{This is a result that demonstrates  that the proposed scheme optimizes the performance of each node individually and also jointly the performance of the complete network.}

Fig.~\ref{fig:energy-vs-snr}(a) depicts the impact of the channel conditions on the energy reduction for $\lambda$=0.5 and \emph{Scenario 1}. This is also an interesting result since it shows that the energy is reduced drastically as the SNR of the RF channel improves. However, at a certain point around 20dB, we see clearly that the energy reduction is minimized. This is because the channel is improved and direct transmissions succeed without requiring retransmissions. Therefore, relaying starts becoming less useful in this case since every node can optimize its own transmissions. On the other hand for poor channel quality, relaying cannot help considerably because all the RF links are affected.

\begin{figure}[t]
\begin{center}
\includegraphics[keepaspectratio,width = 0.5\linewidth]{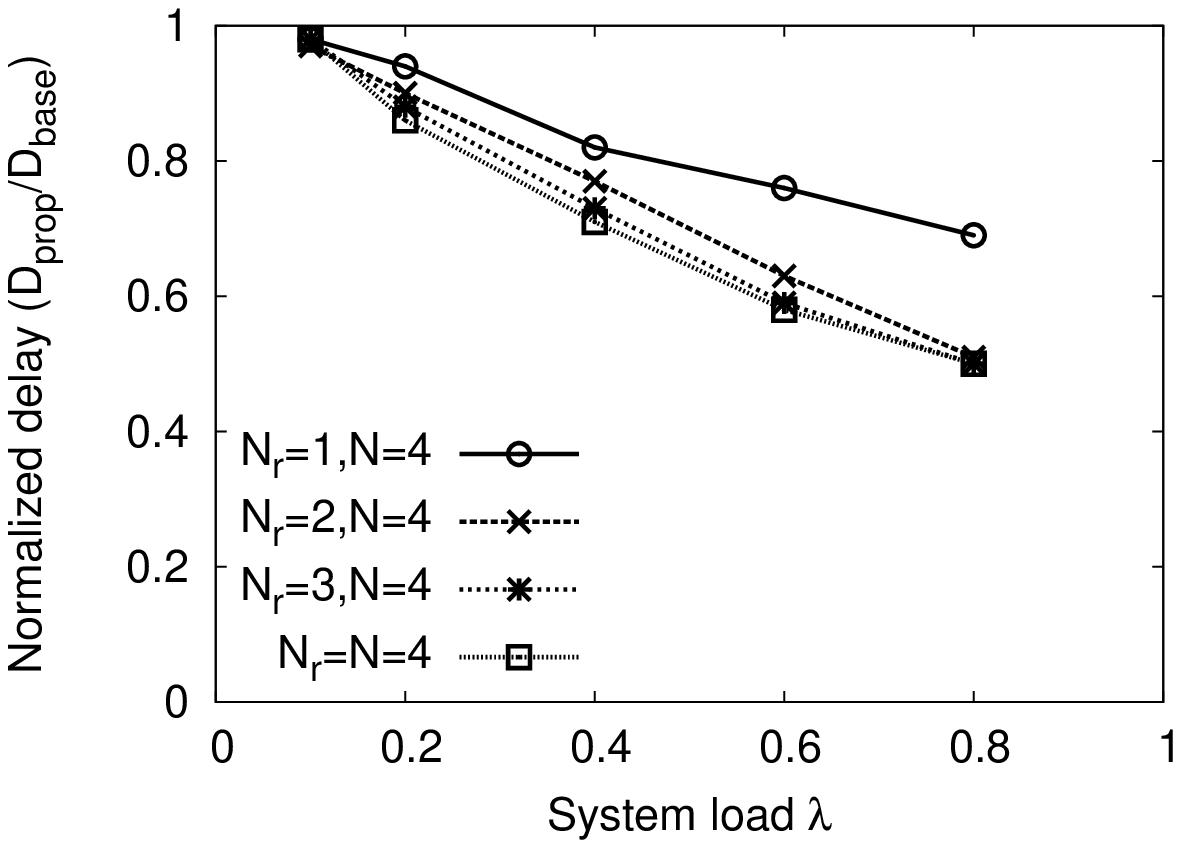}\hspace{-0.2cm}%
\includegraphics[keepaspectratio,width = 0.5\linewidth]{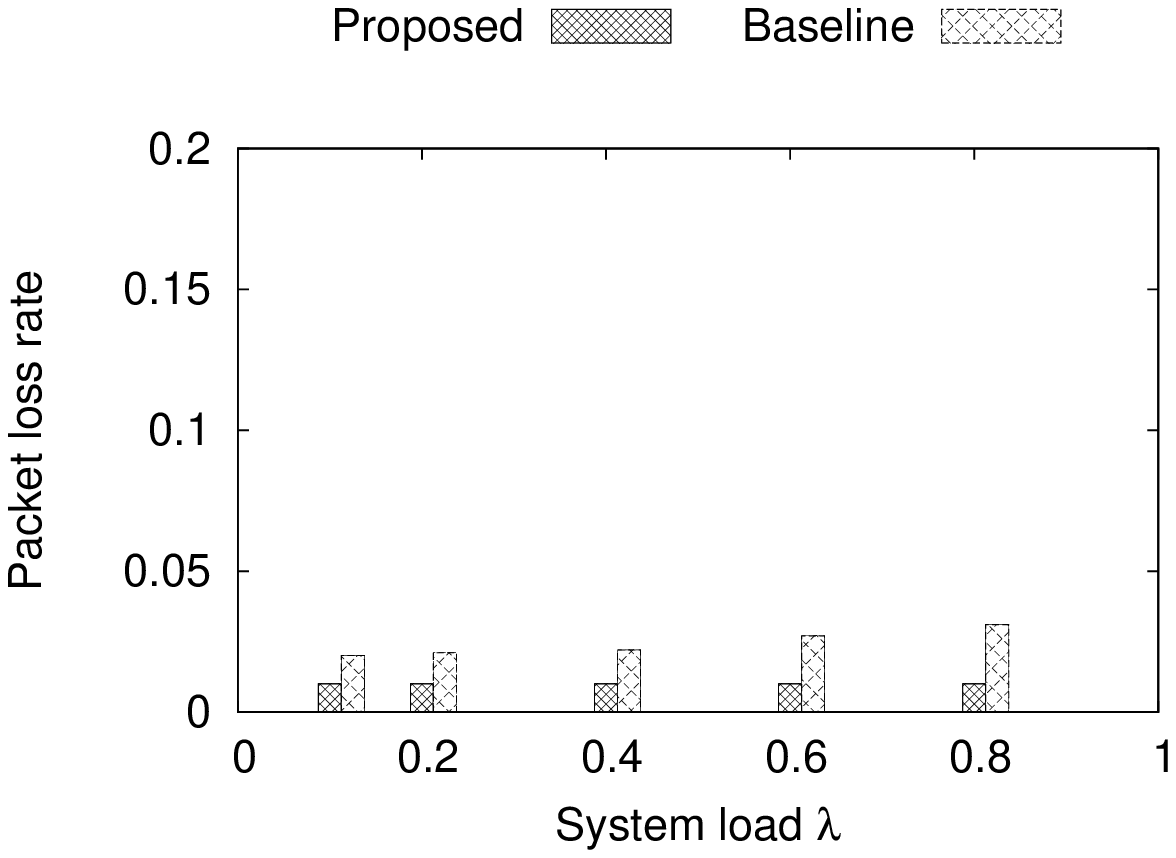}
\caption{Results for the PLR and the relative delay of the proposed system versus the baseline system for $\tau$=1s and \emph{Scenario 1}.}
  \label{fig:latency-vs-load}
\end{center}
\end{figure}

In Fig.~\ref{fig:energy-vs-snr}(b) we present results for $N_r$=2 and $N$=4, while different fixed values of maximum RF transmissions $M_r$, CCA attempts $M_c$, and loads were simulated. Here, $\tau$ was set to 200ms. As the load is increased we observe that there is a need for more CCA attempts than the standard maximum 3 while the value of $M_r$ is not that important for \emph{Scenario 1}. With the proposed scheme and \emph{Scenario 1}, one or at most two transmissions are enough. However, for the cases with $M_r$=1,$M_c$=3, and $M_r$=2,$M_c$=3 of \emph{Scenario 2}, we notice that the ratio is reduced even more. The reason is that for higher $M_r$ the baseline method is very energy inefficient since half of the nodes suffer from high PLR and several retransmissions are needed. On the other hand the proposed system does not use the maximum $M_r$=2 since packets are forwarded through other RF links.

\subsubsection{Results for Delay Minimization}
For this experiment, we reversed the optimization objective and the constraints of problem formulations explained in Section~\ref{section:optimization}. More specifically the objective in this case was delay minimization subject to an energy constraint. The results correspond to different load $\lambda$ and number of relays $N_r$. The total energy constraint was set to 100 Joule that can cover in theory the transmission of 1000 100-byte packets with the selected RF transceiver. For the optimization problem a proportional energy budget was assigned to each packet depending on the remaining energy of the node. The improvement in the average delay and PLR of the transmitted packets can be seen in Fig.~\ref{fig:latency-vs-load}. When every node operates with the baseline system the average number of retransmissions reaches a value between three and four. We observed that this difference is more significant if there is no energy constraint.

\subsection{Results with Real RF Traces}
\begin{figure}[t]
\begin{center}
\subfigure[Delay results ($T_{est}$=1s)]{ \includegraphics[keepaspectratio,width = 0.5\linewidth]{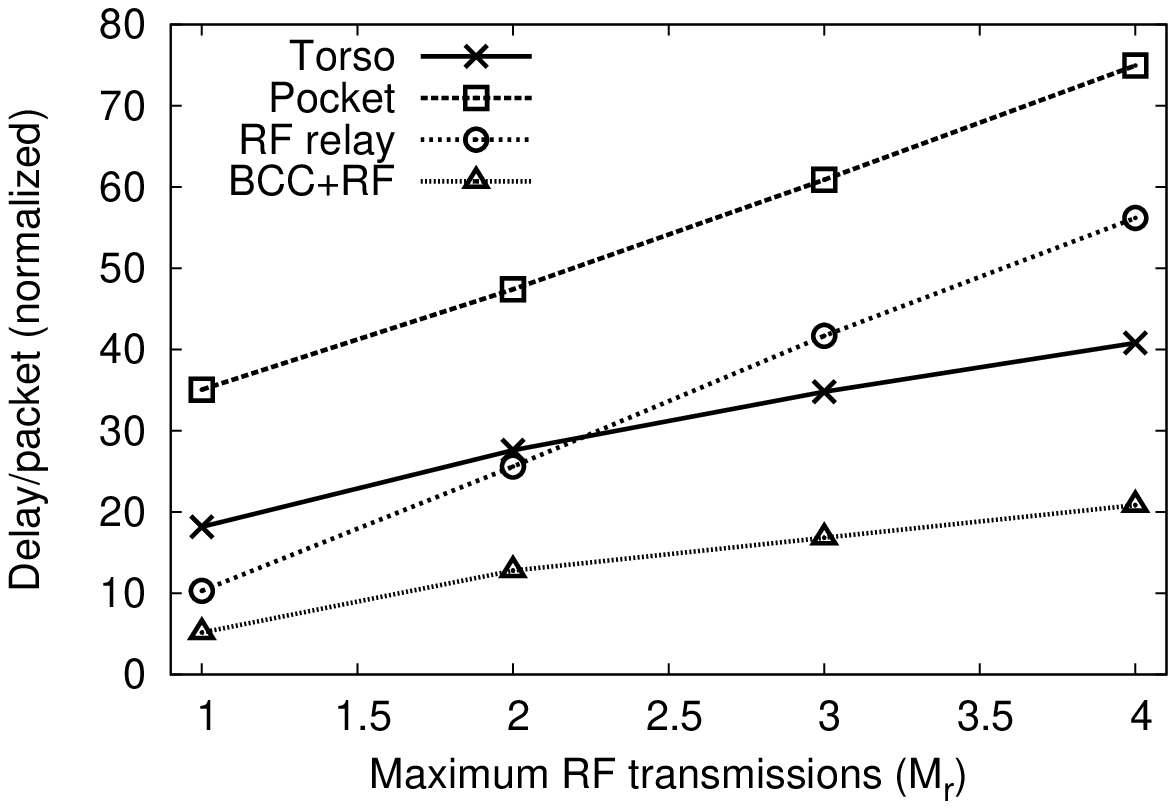}}\hspace{-0.3cm}%
\subfigure[Energy results ($T_{est}$=1s)]{ \includegraphics[keepaspectratio,width = 0.5\linewidth]{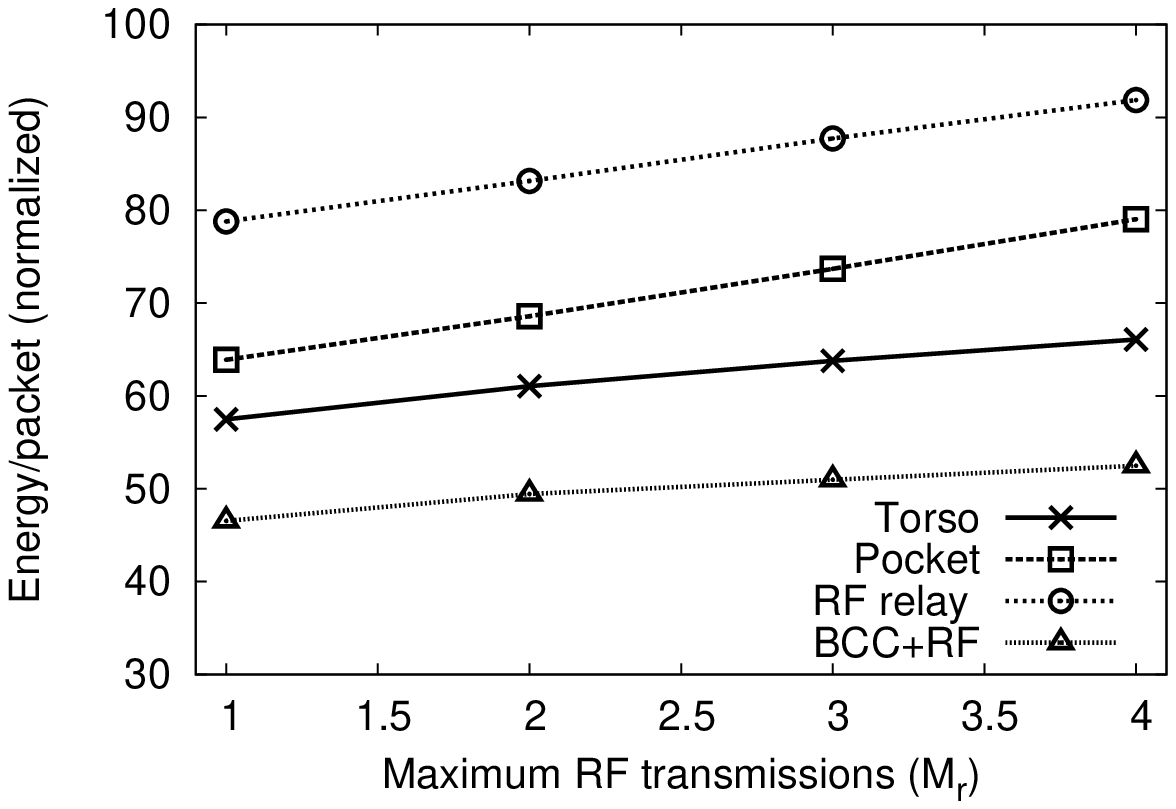}}

\subfigure[PLR results ($T_{est}$=1s)]{ \includegraphics[keepaspectratio,width = 0.5\linewidth]{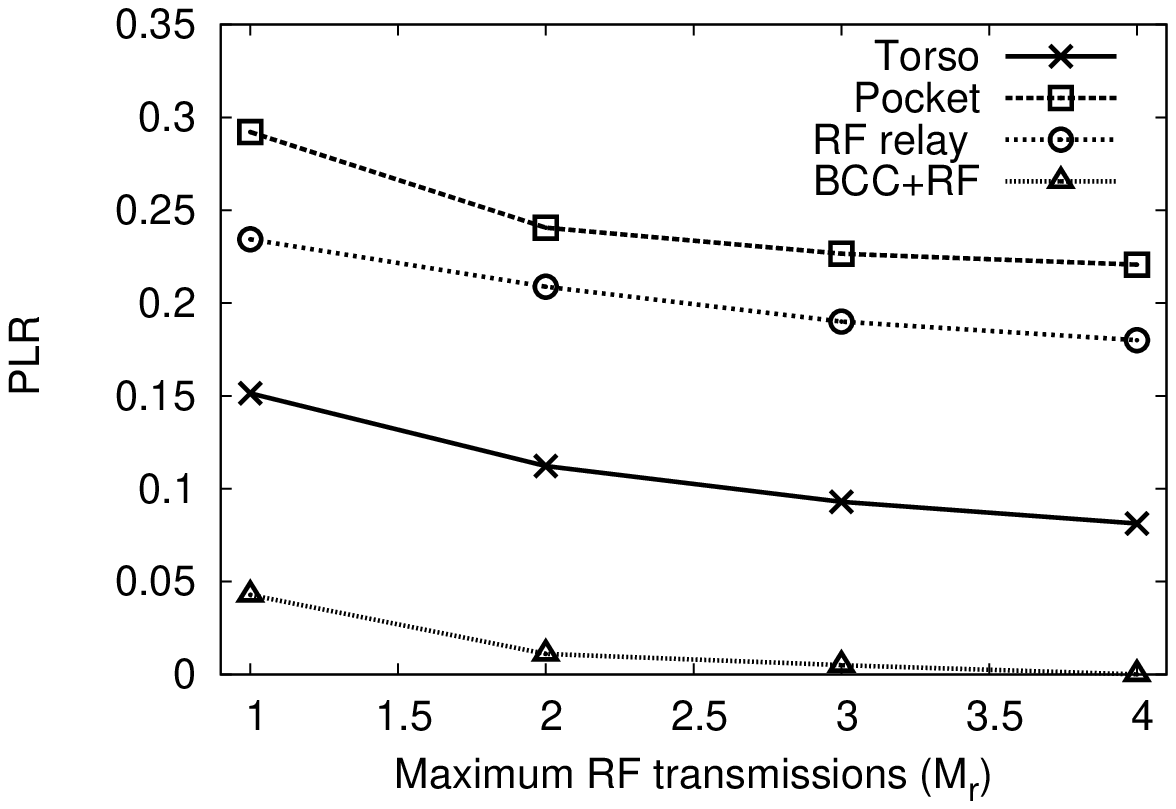}}\hspace{-0.3cm}%
\subfigure[Delay results ($T_{est}$=5s)]{\includegraphics[keepaspectratio,width = 0.5\linewidth]{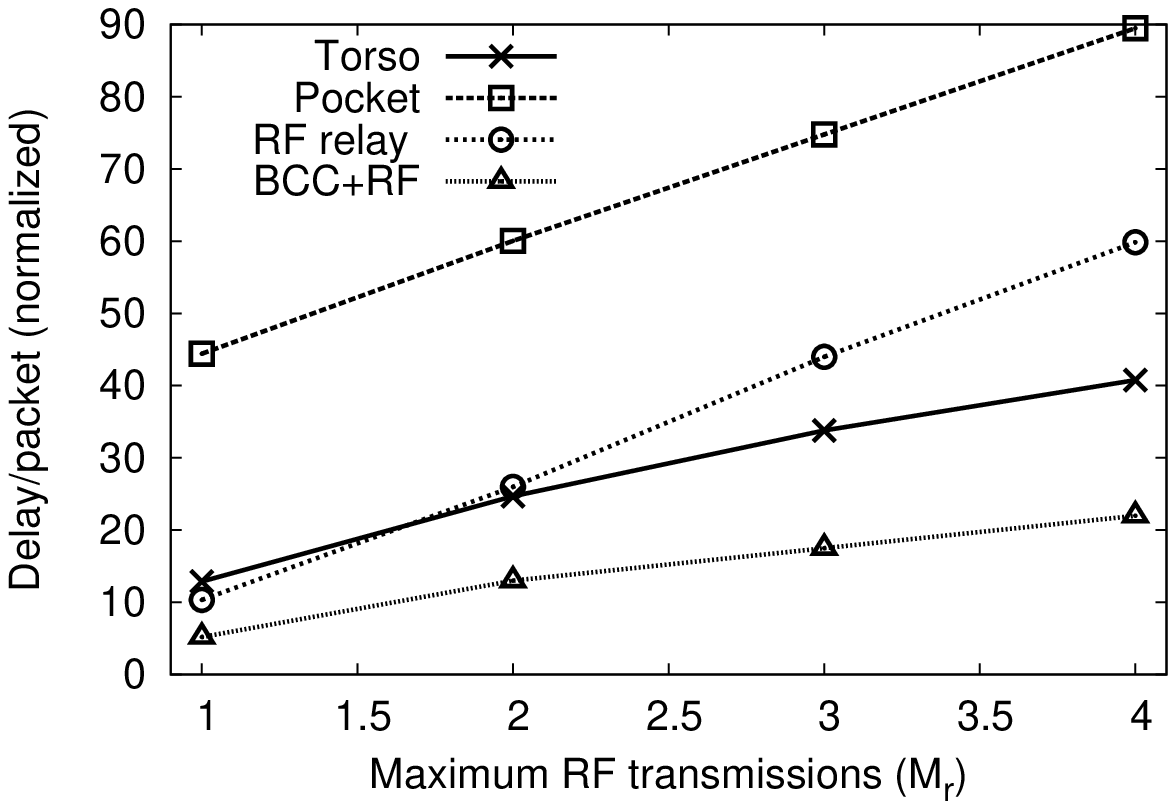}}
\caption{Results for different systems with real RF traces.}
  \label{fig:real-traces}
\end{center}
\end{figure}
\textbf{RF Traces and Setup.} For this part of our evaluation we used the real RSSI traces from two mobile experiments that used the IEEE 802.15.4 and involved two users~\cite{Miluzzo08radiocharacterization}. According to the scenario in~\cite{Miluzzo08radiocharacterization}, a single node is placed in each user and both users walk towards each other starting from a distance of 50m. They continue walking after they cross each other for an additional distance of 10m. In the first experiment both RF nodes are placed in the torso and in the second experiment the node is located in the pocket of a user's trousers. These measurements reflect exactly what we want to test and what~\cite{Miluzzo08radiocharacterization} also showed: for the larger duration of the experiment connectivity between the two torso nodes is very good and better than the experiment where nodes are placed in the pocket. However, when the two users cross each other, connectivity between nodes in the torso is lost completely, while for the experiment with nodes placed in the pocket connectivity is still good. Note also that these traces correspond to low mobility scenarios that is generally the focus of WBAN monitoring applications.

\textbf{Energy Model of the Complete System.} In this case we considered additional sources of power consumption. We include first the power consumption for powering-up/down the RF transceiver as specified in the CC2420 datasheet~\cite{cc2420}. We also approximated the computational overhead of the relay selection algorithm by compiling and profiling in terms of executed instructions our code in the CC2420 simulator. Based on the number of required execution cycles, we used the additional delay and power consumption of the micro-controller in that mode. So in this way we include both communication and computational energy and delay costs.

\textbf{Results.} In our results presented in Fig.~\ref{fig:real-traces} we see the average packet delay for different $M_r$. It is interesting to note that even when both nodes are assumed to be used simultaneously, helping each other as the baseline system we defined earlier (this system here is denoted as "RF relay"), this is not always the optimal case even when compared with just "Torso" and "Pocket". The reason is that there is significant body shadowing between the node in torso and the node in the pocket and so there is significant packet loss that has to be compensated with higher number of retransmissions. Energy and delay results follow a similar trend in Fig.~\ref{fig:real-traces}(b). Regarding the PLR, when $M_r$ is increased this is decreased as Fig.~\ref{fig:real-traces}(c) indicates. The proposed system reaches perfect PLR because the best link is selected and there is no need for many retransmissions.

Finally, we also evaluated the impact of a slower frequency for propagating RF channel estimates. We measured all the previous parameters and we only present delay in Fig.~\ref{fig:real-traces}(d). This higher update frequency mainly affects the "RF relay" system because it makes suboptimal RF relay decisions. Our system is affected with a difference that is barely noticeable. Of course for an even slower update frequency of more than 10 seconds, we observed that similarly the performance of all systems was becoming proportionally worse.

An interesting detail is that the used traces consider backlogged traffic at each node which is inconsistent with the Poisson assumption used in our performance model. Thus, the performance improvement clearly shows that our approach is not limited by the traffic model and it can work under more general settings.

\section{Conclusions}
\label{section:conclusions}
In traditional RF-based WBANs in the face of body shadowing there is a fundamental tradeoff: Either consume more energy for retransmissions or channel coding in order to increase reliability and suffer also higher delay, or reduce delay and energy at the cost of reduced reliability. However, we demonstrated that the above does not have to be the case if nodes cooperate through a delay/energy-efficient secondary link (in this case BCC) in order to select the most efficient RF link for forwarding WBAN data to a gateway. The performance gains are materialized not only with the novel WBAN architecture, but also with a NWK protocol that uses an algorithm for RF relay selection and packet forwarding driven by a performance model. Furthermore, the cross-technology performance models are exploited for local optimization of the MAC parameters of the protocols that operate in the BCC and RF subnetworks. The performance results indicate that the proposed system is more efficient than a state-of-the-art scheme in terms of energy and delay under different realistic channel conditions, application loads, and performance constraints.

The potential concern regarding the proposed system is the need for two different technologies. However, existing IC technology already allows the integration of multiple transceivers in a single chip. Our future work will be focused first on the optimal configuration of other system parameters like the number of used relays and the content of the transmitted packets. Next, we intend to investigate the potential benefits of the proposed system when it is used with an IEEE 802.11 RF communication link for multimedia transmission scenarios.

\section*{Acknowledgment}
The authors would like to acknowledge the anonymous reviewers for their thorough comments that helped improve this manuscript considerably.


\begin{IEEEbiographynophoto}%
{Antonios Argyriou} received the Diploma in electrical and computer engineering from Democritus University of Thrace, Greece, in 2001, and the M.S. and Ph.D. degrees in electrical and computer engineering as a Fulbright scholar from the Georgia Institute of Technology, Atlanta, USA, in 2003 and 2005, respectively.

Currently, he is a tenure-track faculty member at the department of electrical and computer engineering, University of Thessaly, Greece. From 2007 until 2010 he was a Senior Research Scientist at Philips Research, Eindhoven, The Netherlands. From 2004 until 2005, he was a Senior Engineer at Soft.Networks, Atlanta, GA. Dr. Argyriou currently serves in the editorial board of the \textit{Journal of Communications}. He has also served as guest editor for the \textit{IEEE Transactions on Multimedia} Special Issue on Quality-Driven Cross-Layer Design, and he was also a lead guest editor for the \textit{Journal of Communications}, Special Issue on Network Coding and Applications. Dr. Argyriou serves in the TPC of several international conferences and workshops in the area of communications, networking, and signal processing. His current research interests are in the areas of wireless communication systems and networks, and video communications. He is a member of IEEE.
\end{IEEEbiographynophoto}

\vspace*{-2\baselineskip}
\begin{IEEEbiographynophoto}%
{Alberto C. Breva} received the Diploma in electrical and computer engineering from the University of Seville, Spain, in 2008, and the M.S. degree in 2010 from the same University. He was an intern at Philips Research during 2009-2010 where he worked on the topic of body coupled communication.
\end{IEEEbiographynophoto}

\vspace*{-2\baselineskip}
\begin{IEEEbiographynophoto}%
{Marc Aoun}  received his Bachelor degree in Computer and Communications Engineering from the American University of Beirut, Lebanon, in 2004, and his Master’s degree in Communications Engineering from RWTH Aachen, Germany, in 2007. Since 2007 he is a Research Scientist at Philips Research, The Netherlands. His current research interests are in the areas of wireless communication systems, wireless sensor networks, real-time systems and segmentation of large-scale networks.
\end{IEEEbiographynophoto}

\end{document}